\documentclass[twocolumn,aps,prx,superscriptaddress,preprintnumbers]{revtex4-2}

\usepackage{graphicx}
\usepackage{dcolumn}
\usepackage{bm}
\usepackage{lineno, here, physics, tikz, amsmath, xcolor, siunitx}
\usepackage{amssymb}
\usepackage{amsfonts}
\usepackage[
    colorlinks=true,       
    citecolor=blue,        
    linkcolor=blue,        
    urlcolor=blue          
]{hyperref}
\usepackage{cleveref}
\usepackage{booktabs}

\begin{document}


\title{All-optical Implementation of Generalized Quantum Teleportation}

\author{Takaya Hoshi}
\affiliation{OptQC Corp., 3-28-13 Nishi-Ikebukuro, Toshima-ku, Tokyo 171-0021, Japan}

\author{Akito Kawasaki}
\affiliation{OptQC Corp., 3-28-13 Nishi-Ikebukuro, Toshima-ku, Tokyo 171-0021, Japan}

\author{Xiruo Yan}
\affiliation{OptQC Corp., 3-28-13 Nishi-Ikebukuro, Toshima-ku, Tokyo 171-0021, Japan}

\author{Atsushi Sakaguchi}
\affiliation{OptQC Corp., 3-28-13 Nishi-Ikebukuro, Toshima-ku, Tokyo 171-0021, Japan}

\author{Takumi Suzuki}
\affiliation{OptQC Corp., 3-28-13 Nishi-Ikebukuro, Toshima-ku, Tokyo 171-0021, Japan}

\author{Tatsuki Sonoyama}
\affiliation{OptQC Corp., 3-28-13 Nishi-Ikebukuro, Toshima-ku, Tokyo 171-0021, Japan}

\author{Hironari Nagayoshi}
\affiliation{OptQC Corp., 3-28-13 Nishi-Ikebukuro, Toshima-ku, Tokyo 171-0021, Japan}

\author{Kosuke Fukui}
\affiliation{OptQC Corp., 3-28-13 Nishi-Ikebukuro, Toshima-ku, Tokyo 171-0021, Japan}

\author{Kan Takase}
\affiliation{OptQC Corp., 3-28-13 Nishi-Ikebukuro, Toshima-ku, Tokyo 171-0021, Japan}

\author{Warit Asavanant}
\email{warit.asavanant@optqc.com} 
\affiliation{OptQC Corp., 3-28-13 Nishi-Ikebukuro, Toshima-ku, Tokyo 171-0021, Japan}

\date{\today}

\begin{abstract}
Measurement-based continuous-variable optical quantum computing inherently offers high-speed, large-scale operations, yet its practical performance remains constrained by the processing latencies and throughput bottlenecks imposed by classical electronic feedforward circuits. To overcome these limitations, we propose a loss-tolerant, all-optical feedforward (AOFF) architecture for generalized quantum teleportation, which executes programmable single-mode Gaussian operations and realizes arbitrary single-mode linear operations with two successive stages. Quantitative noise analysis under realistic device parameters demonstrates that the architecture suppresses the hardware-induced noise floor to levels potentially compatible with fault-tolerant quantum computing with Gottesman–Kitaev–Preskill qubits over the range of operations required for their Clifford gates. By eliminating optoelectronic conversions, this scheme enables continuous operations with high analog bandwidth and high temporal-mode throughput that drastically reduce circuit runtime. Ultimately, this approach delivers a noise-resilient platform that reconciles operational versatility with the intrinsic speed and bandwidth of optical quantum information processing.
\end{abstract}

\maketitle


\section{Introduction}

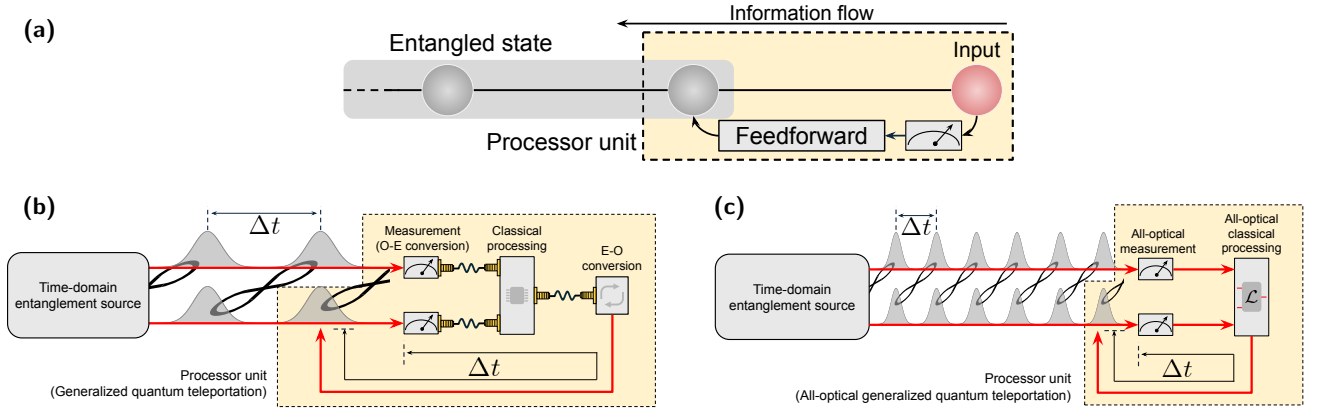
\begin{figure*}[t] 
    \centering
    
    \begin{minipage}{\linewidth}
        \centering
        \begin{tikzpicture}
            \path (0,0) -- (\linewidth,0);
            
            \node[anchor=south, inner sep=0] (image_a) at (0.5\linewidth,0) {\includegraphics[width=0.5\linewidth]{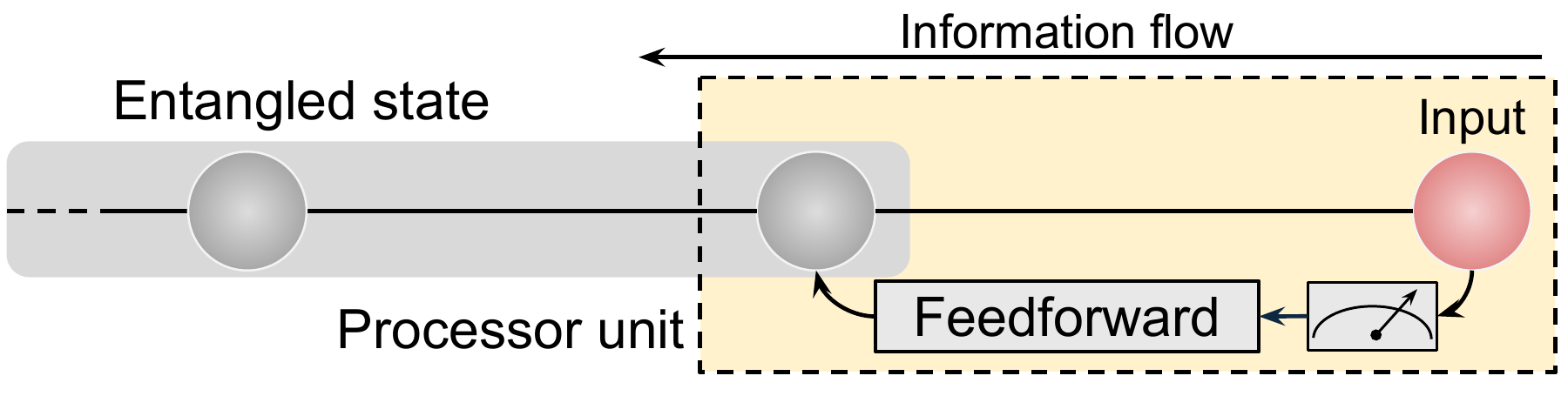}};
            
            \node[anchor=north west, xshift=0.5em, yshift=-0.5em] at (0,0 |- image_a.north) {\textbf{\textsf{(a)}}};
        \end{tikzpicture}
    \end{minipage}

    \vspace{0.6em} 

    \begin{minipage}{0.49\linewidth} 
        \centering
        \begin{tikzpicture}
            \node[anchor=south west, inner sep=0] (image_b) at (0,0) {\includegraphics[width=\linewidth]{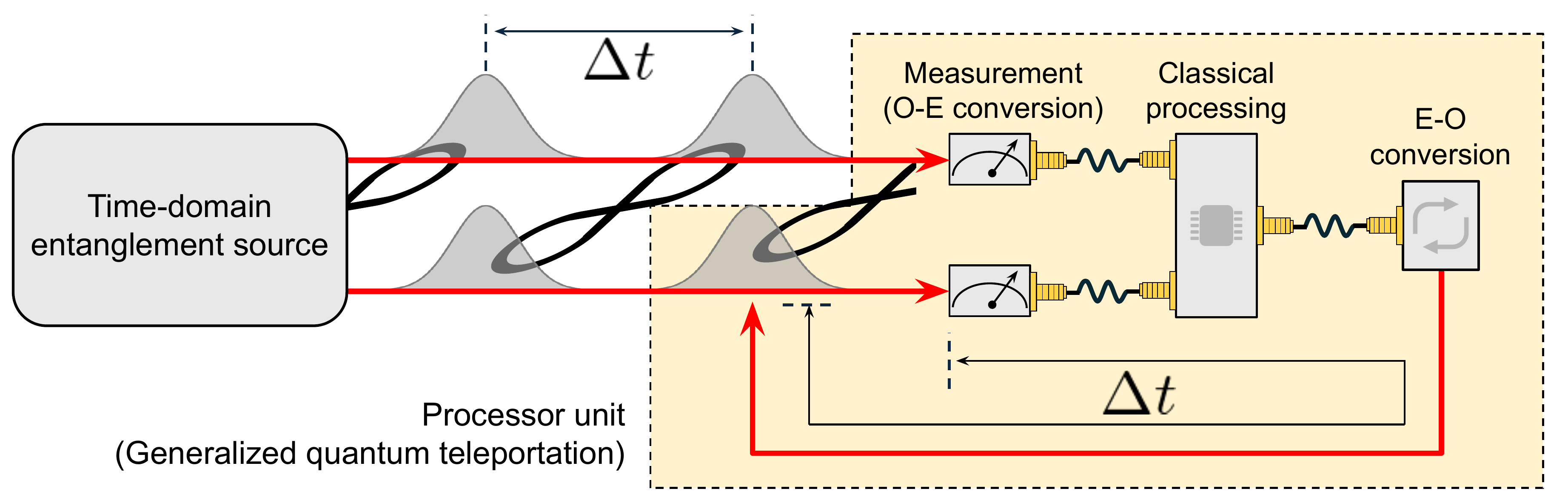}};
            
            \node[anchor=north west, xshift=0.5em, yshift=0.5em] at (image_b.north west) {\textbf{\textsf{(b)}}};
        \end{tikzpicture}
    \end{minipage}
    \hfill 
    \begin{minipage}{0.49\linewidth} 
        \centering
        \begin{tikzpicture}
            \node[anchor=south west, inner sep=0] (image_c) at (0,0) {\includegraphics[width=\linewidth]{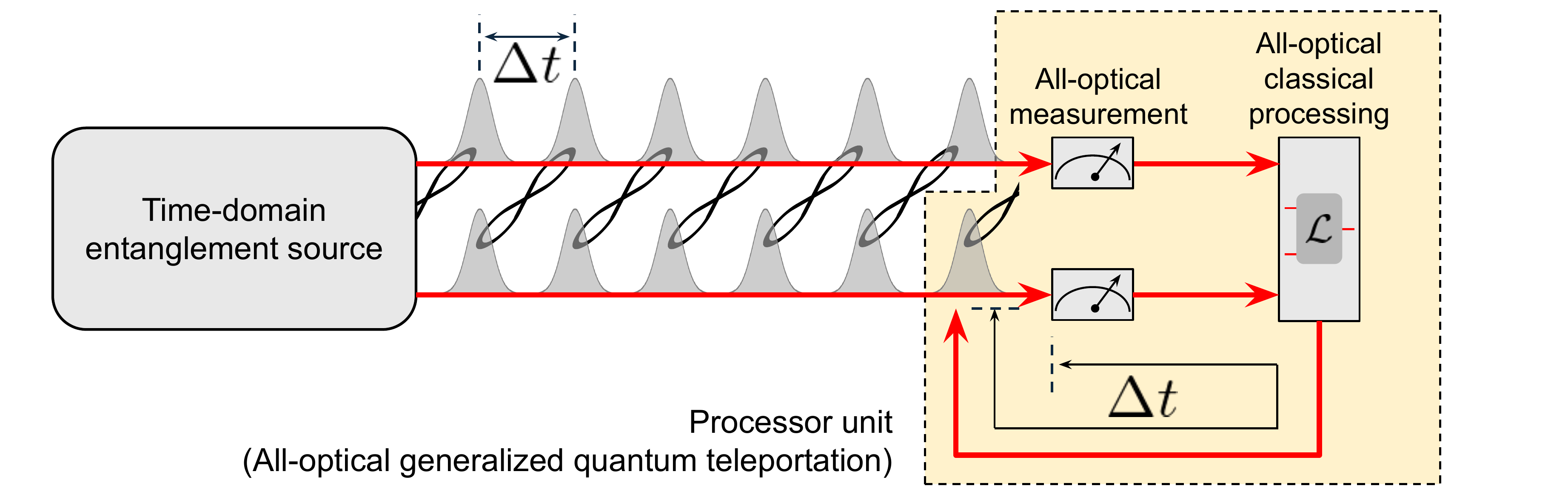}};
            
            \node[anchor=north west, xshift=0.5em, yshift=0.5em] at (image_c.north west) {\textbf{\textsf{(c)}}};
        \end{tikzpicture}
    \end{minipage}
    
    \vspace{0.3em} 

    \caption{Conceptual framework of single-mode measurement-based quantum computing (MBQC) and comparison of feedforward schemes. (a) Conceptual diagram of single-mode MBQC, where the operational clock cycle is determined by measurement and feedforward. (b) Conventional scheme executing feedforward in the electronic domain, requiring a slow optical-to-electronic-to-optical (O-E-O) conversion. (c) Proposed O-E-O-free architecture featuring all-optical feedforward to achieve high throughput and low latency.}
    \label{fig1}
\end{figure*}

While scaling up physical qubit counts remains an indispensable requirement for practical quantum computing platforms, the benchmark for ultimate performance has increasingly expanded to encompass the circuit runtime~\cite{Babbush2021FocusBeyondQuadratic}. In this context, the physical gate execution time itself is not only a vital factor determining performance, but the real-time processing of intermediate measurement outcomes has also attracted intense attention in recent years as another pivotal requirement. Such processing, designed to counteract the intrinsic randomness of quantum measurements to successfully complete a target operation, is collectively referred to as feedforward control~\cite{PhysRevA.71.032318, PRXQuantum.5.010333}. The latency inherent in this real-time process often introduces a severe operational bottleneck, raising profound concerns regarding low-latency classical processing across various domains~\cite{Battistel2023RealTimeDecoding, Caune2026, PRXQuantum.5.010333, 9779088}. This timing constraint becomes critical in measurement-based quantum computing (MBQC), where quantum gates are sequentially constructed through consecutive measurement and feedforward routines~\cite{PhysRevA.71.032318, PhysRevLett.86.5188, PhysRevA.68.022312, PhysRevLett.97.110501, PhysRevA.79.062318}. In MBQC, even the implementation of the most basic quantum operations inherently demands real-time classical processing of measurement outcomes within the feedforward path. As illustrated in Fig.~\ref{fig1}(a), because a single-stage gate operation cannot be finalized until the processed data is fed forward as a correction to the adjacent mode, the hardware latency of even these simple linear classical calculations imposes a strict upper bound on the operational clock rate.

Crucially, mitigating the classical-processing latency bottleneck to increase the operational clock rate inherently requires a correspondingly high temporal processing throughput: as the time interval between consecutive operations is reduced, measurement outcomes must be acquired, processed, and converted into feedforward signals within an increasingly shorter time window. From this perspective, optical quantum computing (OQC) within a time-domain multiplexing (TDM) framework~\cite{PhysRevLett.104.250503, PhysRevA.83.062314, PhysRevA.94.032327, Yokoyama2013UltraLargeScaleCVClusterStates, 10.1063/1.4962732, Asavanant2019TimeDomainMultiplexed2DCluster, Larsen2019Deterministic2DClusterState} natively offers a powerful platform to execute such real-time operations with high throughput. Because optical fields operate at carrier frequencies on the order of hundreds of THz, they provide an enormous bandwidth capable of supporting ultrafast, high-rate operations in the time domain. Within the TDM framework, this broadband capability not only enables exceptionally high throughput but also directly supports scalability; compressing the temporal width of the time slots naturally increases the temporal density of quantum modes per unit time, thereby unifying rapid computation with large-scale scalability. Furthermore, because light propagates as a continuous traveling wave, the sequential execution of such high-density, measurement-based operations is natively accommodated in the time domain. Leveraging this distinct synergy, recent advancements in ultra-broadband squeezed state sources have successfully scaled the source bandwidth into the THz regime~\cite{10.1063/5.0063118,10.1063/1.5142437,suzuki2026broadbandhighlevelsqueezedlight,Chen:22,doi:10.1126/science.abo6213}, thereby paving the way to generate continuous-variable cluster states~\cite{PhysRevLett.97.110501, PhysRevA.79.062318, PhysRevLett.104.250503, PhysRevLett.112.120504, PhysRevA.76.032321, PhysRevA.76.010302, PhysRevLett.98.070502, PhysRevA.78.012301, PhysRevA.83.062314, Yokoyama2013UltraLargeScaleCVClusterStates, 10.1063/1.4962732, PhysRevA.94.032327, PhysRevA.97.032302, Larsen2019Deterministic2DClusterState, Asavanant2019TimeDomainMultiplexed2DCluster, PhysRevResearch.2.023138}---the fundamental resource for TDM-MBQC---within a compact time window~\cite{Hoshi:25}. Nevertheless, despite this massive bandwidth potential of optical fields, state-of-the-art TDM-MBQC systems remain practically constrained to clock rates of up to approximately 100~MHz~\cite{yokoyama2026fullstackanalogopticalquantum,PhysRevApplied.16.034005, Larsen2021}.

This substantial gap between the available resource bandwidth and the actual system speed stems from the underlying computing framework, which relies on generalized quantum teleportation (GQT) as the core mechanism for executing reconfigurable quantum operations~\cite{PhysRevA.81.032315}. To enable programmable quantum gates, the measurement bases must be dynamically tuned across successive time slots [Fig.~\ref{fig1}(b)]~\cite{yokoyama2026fullstackanalogopticalquantum, PhysRevA.93.062326,PhysRevApplied.16.034005, Larsen2021}. Although the classical computation required for even simple operations reduces to a straightforward linear processing of the measurement results, the necessity for base-dependent flexibility has historically forced the framework to rely on electronic-domain detection and subsequent classical electronic circuits. Consequently, the entire feedforward loop encounters severe propagation delays from measurement latency, electronic circuit bandwidth limits, and the inevitable optical-to-electrical-to-optical (O-E-O) conversions, thereby converting simple linear processing into the ultimate operational barrier. While various acceleration techniques have been demonstrated to mitigate these electronic bottlenecks~\cite{Shaked2018, Li:19, PhysRevLett.119.223604, PhysRevA.101.053801, Inoue2023UltraFastOpticalQuantumProcessor, doi:10.1126/science.abo6213, Kawasaki2024BroadbandNonGaussian,Kawasaki2025PicosecondEntanglement, Yamashima2025AllOpticalFeedforward, suzuki2026ultrafastallopticalquantumteleportation,rvmb-ljd3,Kalash:23, Hoshi:25}, most target the analog bandwidth of fixed, non-adaptive operations or isolated components, such as fixed quantum measurement. A scalable, programmable architecture that supports generalized adaptive operations without sacrificing optical bandwidth remains unrealized.

Here, we introduce an all-optical feedforward (AOFF) architecture for GQT. A single GQT stage executes a programmable two-parameter family of single-mode Gaussian operations determined by the measurement angles, and two successive stages realize an arbitrary single-mode linear transformation (see Appendix~\ref{Derivation of generalized quantum teleportation}). As illustrated in Fig.~\ref{fig1}(c), this design replaces the conventional electronic feedforward path with a purely optical routing mechanism, thereby bypassing conversion latencies and electronic bandwidth limits. By operating entirely within the optical domain, this approach preserves the ultra-wide THz analog bandwidth previously restricted to fixed, non-adaptive configurations~\cite{suzuki2026ultrafastallopticalquantumteleportation}. Furthermore, a detailed noise analysis under realistic device parameters demonstrates that this architecture not only enables the programmable execution of these operations but also exhibits high tolerance against optical loss, suppressing the excess noise floor, over the range of operations required for Clifford gates, to levels potentially compatible with fault-tolerant quantum computing (FTQC) platforms with the Gottesman–Kitaev–Preskill (GKP) qubits~\cite{PhysRevA.64.012310, PhysRevX.8.021054, PhysRevA.107.052414}. Collectively, these advancements establish a strong foundation for high-throughput, universal optical quantum computing toward THz-bandwidth operation.

The remainder of this paper is organized as follows. Section~\ref{Preliminary Theory} reviews the preliminary theory of continuous-variable quantum operations and formulates the conventional electronic scheme of GQT. Section~\ref{sec:Proposed_all-optical_FF} presents the proposed all-optical feedforward (AOFF) architecture, detailing its architectural requirements and core operational principles. Section~\ref{sec:Performance_Analysis} quantitatively evaluates the noise performance of both single-stage and multi-stage building blocks under realistic hardware constraints and idealized physical limits. Section~\ref{sec:Discussion} discusses the practical feasibility of integrating this architecture into scalable quantum computing frameworks. This discussion focuses on its compatibility with GKP-based fault-tolerant quantum computing, the engineering requirements for high-speed real-time operation, and remaining future challenges to be addressed for practical implementation. Finally, Section~\ref{conclusion} provides concluding remarks.

\section{Preliminary Theory}\label{Preliminary Theory}

\subsection{Basic quantum operation}

Throughout this study, we adopt the convention $\hbar = 1$. For a single-mode optical field, the annihilation and creation operators are denoted by $\hat{a}$ and $\hat{a}^\dagger$, respectively. The corresponding quadrature operators are defined as $\hat{x} = (\hat{a} + \hat{a}^\dagger)/\sqrt{2}$ and $\hat{p} = (\hat{a} - \hat{a}^\dagger)/(i\sqrt{2})$, which satisfy the canonical commutation relation $[\hat{x}, \hat{p}] = i$.

In the Heisenberg picture, the phase rotation operator $\hat{R}(\theta)$ and its corresponding transformation on the annihilation operator are defined together as:
\begin{equation}\label{eq:rotation-op}
    \hat{R}(\theta) = \exp\left( i\theta\hat{a}^\dagger\hat{a} \right) \quad \Longrightarrow \quad \hat{R}^\dagger(\theta) \hat{a} \hat{R}(\theta) = \hat{a} e^{i\theta}.
\end{equation}

Similarly, the squeezing operator $\hat{S}(r)$ with a squeezing parameter $r \in \mathbb{R}$ and its Bogoliubov transformation are expressed as:
\begin{equation}\label{eq:squeezing-op}
\begin{aligned}
    \hat{S}(r) &= \exp \left[\frac{r}{2}\left(\hat{a}^2 - \hat{a}^{\dagger 2}\right)\right] \quad \\
    &\Longrightarrow \quad \hat{S}^\dagger(r) \hat{a} \hat{S}(r) = \hat{a} \cosh r - \hat{a}^\dagger \sinh r.
\end{aligned}
\end{equation}

For a beam splitter acting on modes $i$ and $j$, we define the splitting angle $\theta$ such that the power transmittance and reflectance are given by $T = \sin^2\theta$ and $R = \cos^2\theta$, respectively. The beam splitter operator $\hat{B}_{ij}(\theta)$ and the resulting linear transformations are defined as follows:
\begin{equation}
\begin{aligned}
    \hat{B}_{ij}(\theta) &= \exp \left[-i \theta\left(\hat{x}_i \hat{p}_j - \hat{p}_i \hat{x}_j\right)\right] \quad \\
    &\Longrightarrow \quad 
    \begin{cases}
        \hat{B}_{ij}^\dagger(\theta) \hat{a}_i \hat{B}_{ij}(\theta) = \hat{a}_i \cos\theta - \hat{a}_j \sin\theta, \\
        \hat{B}_{ij}^\dagger(\theta) \hat{a}_j \hat{B}_{ij}(\theta) = \hat{a}_i \sin\theta + \hat{a}_j \cos\theta.
    \end{cases}
\end{aligned}
\end{equation}

where the graphical circuit notation for this operation is represented as:
\begin{equation*}
    \hat{B}_{ij}(\theta) 
     = 
    \begin{gathered}
    \begin{tikzpicture}[baseline=(current bounding box.center), scale=0.8]
        \definecolor{bsblue}{HTML}{a3c5ed}
        
        \tikzset{
            ray/.style={thick, red},
            bs/.style={draw, fill=bsblue, minimum width=0.7cm, minimum height=0.22cm, inner sep=0pt}
        }
    
        \definecolor{bsblue}{HTML}{a3c5ed}
        \draw[ray] (-1.0, 1.0) -- (1.0, -1.0);
        \draw[ray] (-1.0, -1.0) -- (1.0, 1.0);
    
        \draw[ray] (-1.4, 1.0) -- (-1.0, 1.0);
        \draw[ray] (-1.4, -1.0) -- (-1.0, -1.0);
        \draw[ray] (1.0, 1.0) -- (1.4, 1.0);
        \draw[ray] (1.0, -1.0) -- (1.4, -1.0);
    
        \node[right, font=\small] at (1.4, 1.0) {$i$};
        \node[right, font=\small] at (1.4, -1.0) {$j$};
        \node[left, font=\small] at (-1.4, 1.0) {$i$};
        \node[left, font=\small] at (-1.4, -1.0) {$j$};
    
        \node[bs] at (0,0) {};
    
        \node[right, font=\small] at (0.4, 0.0) {$T$};
    
        \draw[->, >=stealth, thick] (-0.7, 0.5) to[out=225, in=135, looseness=1.2] (-0.7, -0.5);
    
    \end{tikzpicture}
\end{gathered}
\end{equation*}

\subsection{Generalized quantum teleportation}

\begin{figure*}[t]
    \centering
    \includegraphics[width=0.6\linewidth]{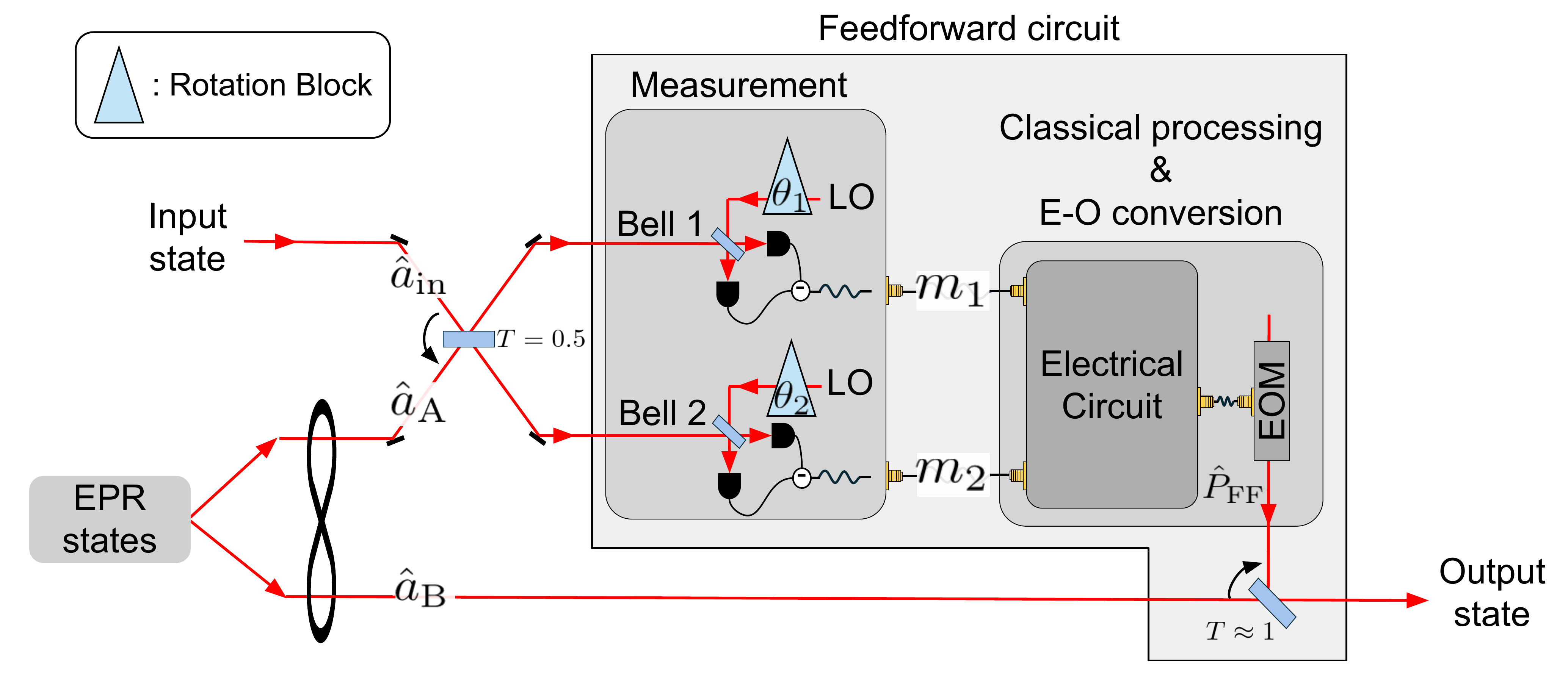}
    \caption{Circuit diagram of generalized quantum teleportation (GQT) with conventional electronic feedforward. The generalized Bell measurement is implemented using a balanced beam splitter followed by homodyne detections, where "Bell 1" and "Bell 2" denote the quantum states after the beam splitter. The measurement outcomes are processed electronically and used to drive electro-optic modulators (EOMs), which realize displacement operations through highly transmissive beam splitters ($T \approx 1$). EPR, Einstein-Podolsky-Rosen; LO, local oscillator.}
    \label{fig2}
\end{figure*}

We first formulate the conventional electronic scheme of GQT to clarify its operational components and terminology. As illustrated in Fig.~\ref{fig2}, the teleportation process begins with a generalized Bell measurement, which is constructed from a balanced beam splitter followed by two homodyne detectors. This setup extends the standard orthogonal measurement to arbitrary, non-orthogonal angles $\theta_1$ and $\theta_2$ ($\theta_1 \neq \theta_2$). In this setup, these measurement angles are physically determined by the phases of the local oscillators (LO) used in the homodyne detections. The input mode $\hat{a}_{\text{in}}$ and mode $\hat{a}_{A}$ of an Einstein-Podolsky-Rosen (EPR) entangled state, which serves as the entangled quantum resource for the teleportation, interfere at a balanced beam splitter, followed by homodyne detection of the two output modes along these generalized angles. The classical electrical signals obtained from the homodyne detection are fed into the classical processing unit, which processes the measurement results according to the target quantum operation to calculate the required displacement amplitudes. These processed signals are then used to drive the electro-optic modulators (EOMs), converting the electronic data back into the optical domain to execute the final displacement operations on the remaining EPR mode $\hat{a}_{B}$. Throughout this paper, the entire sequence spanning from the homodyne detection to these final displacement operations is defined as the feedforward circuit.

Specifically, during the generalized Bell measurement, the homodyne detectors measure the rotated quadratures defined by the transformation $\hat{R}^\dagger(\theta_k) \hat{p} \hat{R}(\theta_k) = \hat{p}\cos\theta_k + \hat{x}\sin\theta_k$ for $k = 1, 2$. The measured operators are given by:

\begin{equation}\label{bell-measurement1}
\begin{aligned}
    \begin{pmatrix}
    \hat{m}_1 \\
    \hat{m}_2
    \end{pmatrix} &= 
    \frac{1}{\sqrt{2}}\begin{pmatrix}
        \sin\theta_1 & \cos\theta_1 \\
        \sin\theta_2 & \cos\theta_2
    \end{pmatrix}\begin{pmatrix}
    \hat{x}_{\text{in}} \\
    \hat{p}_{\text{in}}
    \end{pmatrix} \\
    &+ \frac{1}{\sqrt{2}}\begin{pmatrix}
    -\sin\theta_1 & -\cos\theta_1 \\
    \sin\theta_2 & \cos\theta_2
    \end{pmatrix}\begin{pmatrix}
    \hat{x}_{A} \\
    \hat{p}_{A}
    \end{pmatrix}.
\end{aligned}
\end{equation}

For notational simplicity, the hats are omitted hereafter when referring to the actual experimental outcomes (c-numbers) $m_1$ and $m_2$.

To physically execute the target quantum operation, the implementation of a feedforward operation is indispensable. Specifically, the measurement outcomes $m_1$ and $m_2$ are utilized as classical feedforward signals to drive the final displacement correction. This displacement operation is implemented by coupling mode $\hat{a}_B$ with the feedforward field $\hat{P}_{\text{FF}} = \frac{1}{\sqrt{2}}(\mathcal{X}_{\text{FF}} + i\mathcal{P}_{\text{FF}}) + \hat{a}_{\text{vac}}$ at the displacement beam splitter with power transmittance $T$, where the displacements $\mathcal{X}_{\text{FF}}$ and $\mathcal{P}_{\text{FF}}$ are determined as functions of $m_1$ and $m_2$, and $\hat{a}_{\text{vac}}$ denotes the vacuum fluctuations accompanying the feedforward field. At this beam splitter, mode $\hat{a}_B$ is scaled by a factor of $\sqrt{T}$, whereas the feedforward field is scaled by a factor of $\sqrt{1-T}$. The quadratures of the output mode $\hat{a}_{\text{out}}$ are thus obtained as (see Appendix~\ref{Derivation of generalized quantum teleportation} for details):
\begin{widetext}
\begin{equation}\label{output-GQT}
    \begin{aligned}
    \begin{pmatrix} \hat{x}_{\text{out}} \\ \hat{p}_{\text{out}} \end{pmatrix}
    &= \sqrt{T}\left[\begin{pmatrix} \hat{x}_{B} - \hat{x}_A \\ \hat{p}_{B} + \hat{p}_A \end{pmatrix}
    + \mathbf{V}(\theta_1, \theta_2) \begin{pmatrix} \hat{x}_{\text{in}} \\ \hat{p}_{\text{in}} \end{pmatrix}
    + \frac{\sqrt{2}}{\sin (\theta_2-\theta_1)}
    \begin{pmatrix} m_1\cos \theta_2 + m_2\cos \theta_1 \\ m_1\sin \theta_2 + m_2\sin \theta_1 \end{pmatrix} \right]
    \\
    &+ \sqrt{1-T} \begin{pmatrix} \mathcal{X}_{\text{FF}} \\ \mathcal{P}_{\text{FF}} \end{pmatrix}
    + \sqrt{1-T} \begin{pmatrix} \hat{x}_{\text{vac}} \\ \hat{p}_{\text{vac}} \end{pmatrix},
    \end{aligned}
\end{equation}
\end{widetext}
where $\mathbf{V}(\theta_1, \theta_2)$ represents the Gaussian operation determined by the choice of the measurement angles $\theta_1$ and $\theta_2$. Inside the brackets, the first term corresponds to the quantum noise of the EPR resource state, which vanishes in the infinite squeezing limit, the second term is the target transformation of the input, and the third term is the measurement-induced random displacement. The overall factor $\sqrt{T}$ represents the deterministic amplitude attenuation of mode $\hat{a}_B$ at the displacement beam splitter. In practice, by employing a high-transmittance beam splitter ($T \to 1$), this effect can be made arbitrarily small; for instance, $T = 0.99$ and $T = 0.999$ yield an amplitude attenuation of only $\approx 0.5\%$ and $\approx 0.05\%$ (corresponding to power losses of $1\%$ and $0.1\%$), respectively. To perfectly cancel the measurement-induced random displacement (the third term in the brackets, which is also attenuated by $\sqrt{T}$), the feedforward gain is calibrated to satisfy:
\begin{equation}\label{ideal-feedfoward}
\begin{aligned}
    \hat{P}_{\text{FF}} &= \frac{1}{\sqrt{2}}\left( \mathcal{X}_{\text{FF}} + i\mathcal{P}_{\text{FF}} \right) + \hat{a}_{\text{vac}} \\
    &= e^{i\pi\mathbf{H}(\theta_2-\theta_1)}\sqrt{\frac{T}{1-T}}\\
    &\cdot\frac{\sqrt{2}}{|\sin (\theta_2-\theta_1)|}\left[ \frac{1}{\sqrt{2}}\left\{ m_1 e^{i\theta_2} + m_2 e^{i\theta_1} \right\}\right] + \hat{a}_{\text{vac}},
\end{aligned}
\end{equation}
where the Heaviside step function $\mathbf{H}(\cdot)$ accounts for the phase flip determined by the sign of the measurement basis difference. Substituting Eq.~\eqref{ideal-feedfoward} into Eq.~\eqref{output-GQT} eliminates the measurement-dependent terms, yielding the target transformed state $\sqrt{T}\,\mathbf{V}(\theta_1,\theta_2)(\hat{x}_{\text{in}},\hat{p}_{\text{in}})^{\rm T}$ accompanied by the EPR resource noise and the vacuum contribution $\sqrt{1-T}(\hat{x}_{\text{vac}},\hat{p}_{\text{vac}})^{\rm T}$. 
In the all-optical implementation proposed below, the feedforward field contains no such independent vacuum contribution; instead, the de-amplified quadratures of the phase-sensitive amplification (PSA), which are canonically conjugate to the amplified ones, ensure the commutation relation of the feedforward mode (see Appendix~\ref{Residual Noise Contributions from Non-Amplified Quadratures}).

\section{Proposed all-optical feedforward architecture}\label{sec:Proposed_all-optical_FF}

\subsection{Requirements on the Proposed Architecture}\label{Requirements on the Proposed Architecture}

To fully exploit the intrinsic bandwidth and scalability of continuous-variable optical fields under the TDM-MBQC framework, the realization of high-throughput feedforward operations is indispensable. Fulfilling this condition demands a processing mechanism that operates entirely within the optical domain. Shifting the processing completely into the optical domain, however, inherently introduces additional technical challenges: ensuring robustness against optical losses and mitigating physical circuit complexity. Therefore, a viable architectural framework must be systematically designed based on the following specific criteria to overcome these accompanying hurdles while preserving the ultimate benefits of its all-optical nature.

First, the high-throughput requirement dictates that the feedforward operations be executed without the speed bottlenecks of classical electronics. In conventional configurations, the electronic feedforward loop inherently restricts the clock rate to 100~MHz \cite{yokoyama2026fullstackanalogopticalquantum} due to the propagation delays of optical-to-electrical-to-optical (O-E-O) conversions and the finite throughput of classical circuits. Overcoming this bottleneck necessitates bypassing intermediate homodyne detection, classical electronic processing, and electro-optic modulation, thereby executing the entire feedforward process purely within the optical domain to allow ultrafast operation.

Second, loss robustness is strictly demanded to preserve quantum information while sustaining this high operational speed, with the ultimate goal of rendering the entire feedforward process fundamentally lossless. In an all-optical routing configuration, any optical losses encountered along the paths from the measurement stage to the final displacement operation can easily degrade the quantum signals. To counteract these losses and approach an ideal, lossless operation, the system requires an optical processing mechanism that can amplify the signal before attenuation occurs, without introducing any severe noise penalties. To achieve this, the framework must selectively extract and amplify the target quadrature component entirely within the optical domain, serving as the optical analogue to conventional homodyne detection. Integrating a PSA mechanism \cite{Ralph:99, Yamashima2025AllOpticalFeedforward, suzuki2026ultrafastallopticalquantumteleportation} is essential to fulfilling this requirement. Because a PSA executes directional amplification along a specific phase axis, it can ideally amplify the target quadrature without introducing any additional quantum noise. Furthermore, operating this noise-free amplification in a high-gain regime is the key to unlocking complete loss robustness. Providing a sufficiently high amplification gain lifts the extracted quadrature information far above the vacuum fluctuation level, rendering the quantum signals exceptionally robust against any subsequent downstream losses. Ultimately, assuming an ideal PSA characterized by zero internal loss, this combination of noiseless operation and high amplification gain theoretically allows the entire feedforward network to approach a completely lossless operation, successfully preserving a high signal-to-noise ratio.

Third, the architecture must exhibit structural simplicity to avoid excessive hardware overhead. While the required feedforward relations involve mathematically complex transformations depending on the measurement angles, the physical network should execute these operations through a streamlined configuration.

\subsection{Operational Principle and Derivation}\label{Operational Principle and Derivation}

\begin{figure*}[htbp]
    \centering
    \includegraphics[width=0.7\linewidth]{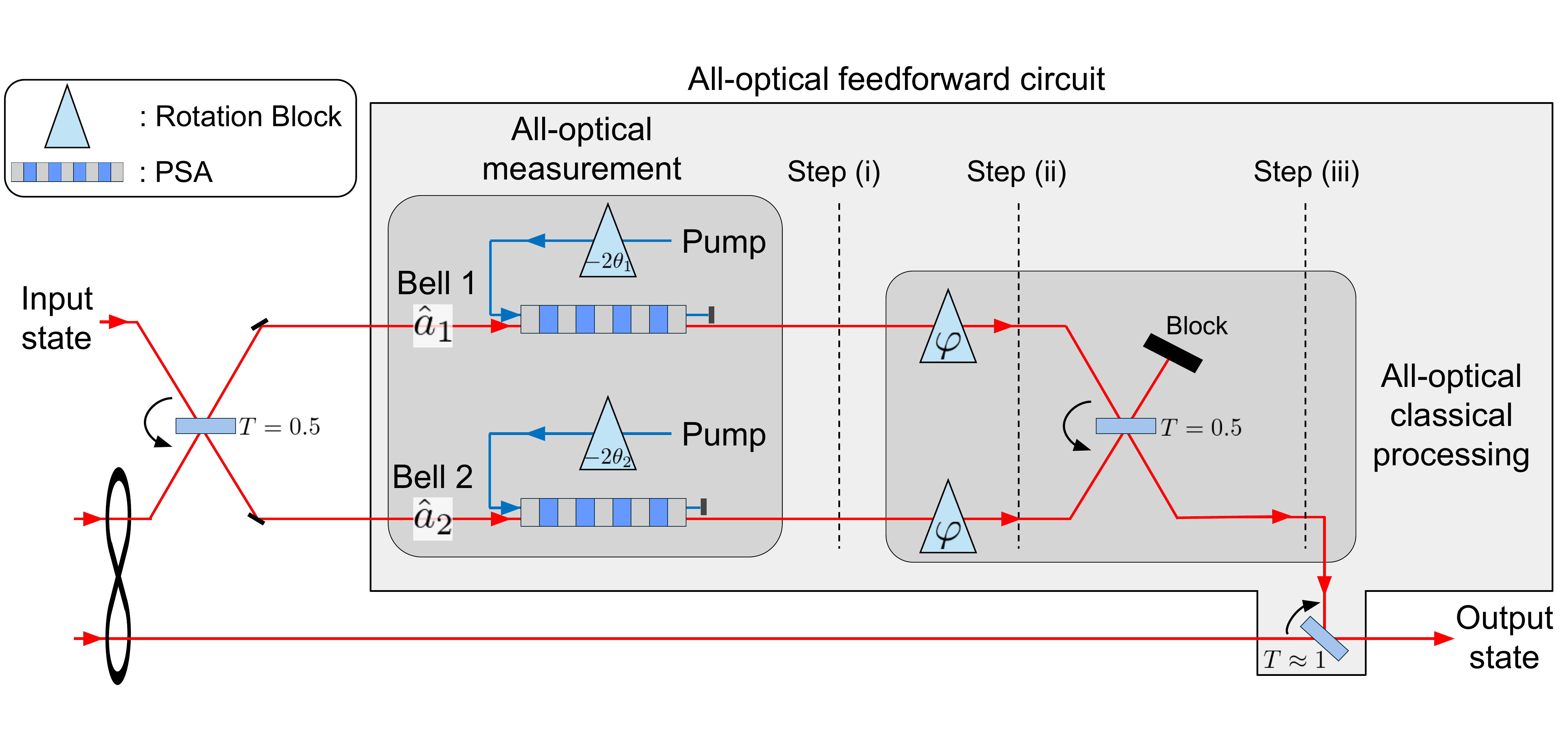}
    \caption{Overall circuit for all-optical generalized quantum teleportation (GQT), highlighting the core all-optical feedforward (AOFF) circuit that integrates all-optical measurement and optical-domain classical signal processing using phase-sensitive amplifiers (PSAs).}
    \label{fig3}
\end{figure*}

Figure \ref{fig3} illustrates the concrete hardware configuration designed to simultaneously satisfy all the design criteria established in Sec.~\ref{Requirements on the Proposed Architecture}: high throughput, loss robustness, and structural simplicity. In this subsection, we elaborate on the operational principles of this proposed AOFF architecture and present the rigorous mathematical derivation of the target transformations. The AOFF operation proceeds via a three-step protocol:

Step (i): Information Extraction. Here, the PSAs are utilized to optically extract the target quadrature information from the Bell-measurement modes $\hat{a}_k$ ($k=1,2$). To describe this process, we introduce the rotated quadrature operators $\hat{x}_k(\theta_k)$ and $\hat{p}_k(\theta_k)$ defined along the measurement angle $\theta_k$. These operators are generated as $\hat{x}_k(\theta_k) = \hat{R}^\dagger(\theta_k) \hat{x}_k \hat{R}(\theta_k)$ and $\hat{p}_k(\theta_k) = \hat{R}^\dagger(\theta_k) \hat{p}_k \hat{R}(\theta_k)$, which decompose the annihilation operator as $\hat{a}_k = \frac{1}{\sqrt{2}} e^{-i\theta_k} (\hat{x}_k(\theta_k) + i\hat{p}_k(\theta_k))$. By matching the pump phase of the optical parametric amplifier to $-2\theta_k$, the PSA selectively amplifies the target quadrature component $\hat{m}_k \equiv \hat{p}_k(\theta_k)$ while simultaneously de-amplifying (attenuating) the orthogonal component $\hat{x}_k(\theta_k)$. This phase-sensitive mechanism allows the circuit to extract the necessary quadrature information while discarding the unwanted orthogonal information. Described by the transformation operator $\hat{S}_{\theta_k}(r) = \hat{R}^\dagger(\theta_k)\hat{S}(r)\hat{R}(\theta_k)$, where the explicit argument $r$ is hereafter omitted for brevity (see Appendix~\ref{Decomposition and Phase Control of Phase-Sensitive Amplification}), this process yields:
\begin{equation}\label{eq:measop}
    \hat{S}_{\theta_k}^\dagger \hat{a}_k \hat{S}_{\theta_k} = \frac{1}{\sqrt{2}} e^{-i\theta_k} (e^{-r} \hat{x}_k(\theta_k) + i e^r \hat{p}_k(\theta_k)).
\end{equation}

In the high-gain limit ($G = e^{2r} \gg 1$), the amplified quadrature component heavily dominates, whereas the de-amplified orthogonal component becomes negligibly small and is effectively discarded. Substituting $i = e^{i\pi/2}$, the output mode is tightly approximated as:
\begin{equation}\label{stepi-AOLFF}
    \hat{a}_{k,\text{out}} = \hat{S}_{\theta_k}^\dagger \hat{a}_k \hat{S}_{\theta_k} \approx \frac{e^{i(\pi/2 - \theta_k)}}{\sqrt{2}} \sqrt{G} \hat{m}_k.
\end{equation}
While this approximation captures the primary dynamics of the AOFF circuit, the de-amplified components remain at finite gain. In Appendix~\ref{Residual Noise Contributions from Non-Amplified Quadratures}, we derive the exact input--output relation retaining these components and show that they only induce a deterministic modification of the implemented Gaussian operation of order $1/G$, together with a small input-independent term originating from the EPR resource.

Step (ii): State Transformation. To align the quadratures in phase space, a phase rotation $\hat{R}(\varphi)$ with $\varphi = -\pi/2 + \theta_1 + \theta_2$ is applied to each of the two measurement channels. The resulting mode transformation is expressed as:
\begin{equation}\label{stepii-AOLFF}
    \begin{aligned}
        \hat{a}_{1}' &= \hat{S}_{\theta_1}^\dagger \hat{R}^\dagger(\varphi)\hat{a}_1 \hat{R}(\varphi)\hat{S}_{\theta_1} \\
        &= e^{i\varphi}\hat{a}_{1,\text{out}} \\
        &= e^{i(-\frac{\pi}{2} + \theta_1 + \theta_2)} \cdot \frac{e^{i(\frac{\pi}{2} - \theta_1)}}{\sqrt{2}} \sqrt{G} \hat{m}_1 \\
        &= \frac{e^{i\theta_2}}{\sqrt{2}} \sqrt{G} \hat{m}_1, \\
        \hat{a}_{2}' &= \hat{S}_{\theta_2}^\dagger \hat{R}^\dagger(\varphi)\hat{a}_2 \hat{R}(\varphi)\hat{S}_{\theta_2} \\
        &= e^{i\varphi}\hat{a}_{2,\text{out}} \\
        &= e^{i(-\frac{\pi}{2} + \theta_1 + \theta_2)} \cdot \frac{e^{i(\frac{\pi}{2} - \theta_2)}}{\sqrt{2}} \sqrt{G} \hat{m}_2 \\
        &= \frac{e^{i\theta_1}}{\sqrt{2}} \sqrt{G} \hat{m}_2.
    \end{aligned}
\end{equation}
The $\pi$ phase shift associated with the Heaviside step function $\mathbf{H}(\theta_2 - \theta_1)$ is incorporated by shifting the phase to $\varphi \to \varphi + \pi$.

Step (iii): Displacement Signal Generation. To generate the all-optical feedforward signal, the two rotated modes are interfered at a balanced beam splitter:
\begin{equation}\label{measinfo-AO}
    \hat{P}_{\text{FF, AO}} = \frac{1}{\sqrt{2}} (\hat{a}_1' + \hat{a}_2') = \frac{\sqrt{G}}{2} \left( \hat{m}_1 e^{i\theta_2} + \hat{m}_2 e^{i\theta_1} \right).
\end{equation}
Comparing Eq.~\eqref{measinfo-AO} with the ideal feedforward relation in Eq.~\eqref{ideal-feedfoward}, the required parametric gain is determined by:
\begin{equation}\label{required-gain}
\sqrt{G} = \frac{2\sqrt{T}}{\sqrt{1-T}|\sin(\theta_2 - \theta_1)|}.
\end{equation}
Here, the factor $\sqrt{T}$ in the numerator reflects the attenuation of the measurement-induced displacement of mode $\hat{a}_B$ at the displacement beam splitter [Eq.~\eqref{output-GQT}]. This expression assumes an ideal lossless system; in practical implementations, $G$ acts as an effective gain that must be adjusted to compensate for physical imperfections such as optical losses, as analyzed in Sec.~\ref{sec:Performance_Analysis}. Note that the phases, gain, and beam-splitter ratios are assumed to be calibrated exactly here; the effects of calibration errors are discussed in Appendix~\ref{Residual Noise Contributions from Non-Amplified Quadratures}.

\section{Performance Analysis}
\label{sec:Performance_Analysis}

Based on the principles described above, we quantitatively analyze the performance of the AOFF architecture. By incorporating realistic imperfections into each physical component, we evaluate the overall noise characteristics and quantify the excess noise introduced by finite amplifier gain and optical losses.

\subsection{Framework for Covariance Matrix Propagation}

The noise performance of the proposed architecture is evaluated using the covariance matrix formalism~\cite{RevModPhys.84.621}. For an $n$-mode quantum state, the quadrature vector is defined as $\hat{\boldsymbol{r}} = (\hat{x}_1, \hat{p}_1, \dots, \hat{x}_n, \hat{p}_n)^T$, with the elements of the covariance matrix $V$ given by $V_{ij} = \frac{1}{2} \langle \{ \hat{r}_i, \hat{r}_j \} \rangle - \langle \hat{r}_i \rangle \langle \hat{r}_j \rangle$. In the Heisenberg picture, we model the entire network as a sequence of discrete optical processing steps. The $l$-th step, which represents a specific physical process such as squeezing, phase rotation, or optical loss, evolves the quadratures according to:
\begin{equation}\label{covariancematrix-rule}
\hat{\boldsymbol{r}}_{l+1} = W_l \hat{\boldsymbol{r}}_l + \hat{\boldsymbol{\xi}}_l,
\end{equation}
where $W_l$ is the transformation matrix of the $l$-th step and $\hat{\boldsymbol{\xi}}_l$ is the associated Gaussian noise vector. Assuming the added noise is independent of the input state, the covariance matrix evolves via the Gaussian channel:
\begin{equation}\label{cov-evolution}
V_{l+1} = W_l V_l W_l^T + V_{\xi,l},
\end{equation}
where $V_{\xi,l} = \frac{1}{2} \langle \{ \Delta\hat{\boldsymbol{\xi}}_l, \Delta\hat{\boldsymbol{\xi}}_l^{T} \} \rangle$ is the covariance matrix of the noise introduced at step $l$.

Recursive application of Eq.~\eqref{cov-evolution} across a sequence of $L$ total steps yields the total accumulated noise $V_{\text{noise}, L}$ at the final output:
\begin{align}\label{general-noise-sum}
V_{\text{noise}, L} &= \sum_{i=0}^{L-1} \mathbf{M}_{i \to L} V_{\xi, i} \mathbf{M}_{i \to L}^T, \\
\text{where} \quad \mathbf{M}_{i \to L} &= 
\begin{cases}
W_{L-1} W_{L-2} \cdots W_{i+1} & (i < L-1), \\
\mathbf{I} & (i = L-1).
\end{cases} \nonumber
\end{align}
This framework accounts for the propagation of the noise generated at each individual step through all subsequent optical components to the output.

\subsection{Noise Evaluation of the Single-Stage Building Block}

\begin{figure*}[!ht]
    \centering
    \includegraphics[width=0.7\linewidth]{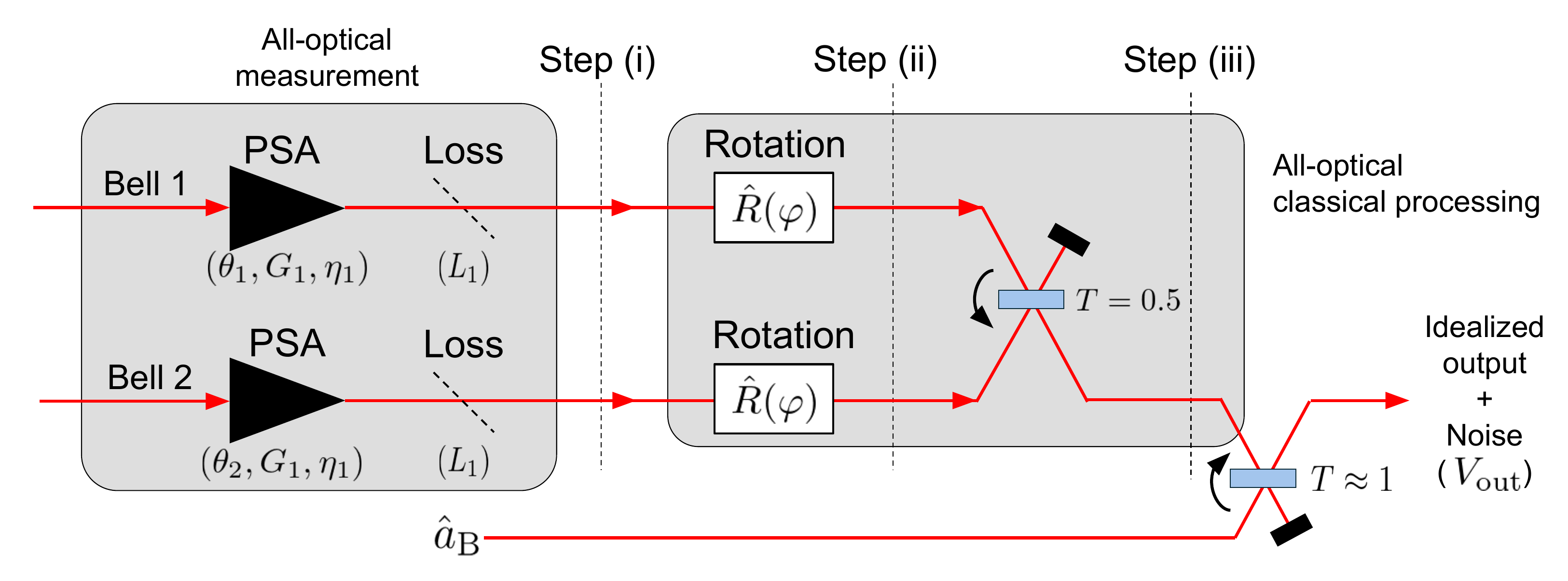}
    \caption{Mathematical modeling of the single-stage building block (Fig.~\ref{fig3}) incorporating experimental imperfections. The setup parameterizes the non-ideal PSA (gain $G_1$, internal efficiency $\eta_1$, phase $\theta_k$), external propagation loss $L_1$, and feedforward phase rotation $\varphi$. It further includes a 50:50 BS and a high-transmittance BS ($T \sim 1$) for displacement coupling with mode $\hat{a}_B$, yielding the final output noise covariance matrix $V_{\text{out}}$ [Eq.~\eqref{single-stage-total-noise}].}
    \label{fig4}
\end{figure*}

In this subsection, we evaluate the noise performance of the single-stage building block shown in Fig.~\ref{fig3} under realistic experimental imperfections. As stated in Sec.~\ref{Operational Principle and Derivation}, the phases, gain calibration, and beam-splitter ratios are assumed to be exact, so that the analysis below focuses on the additive noise induced by the hardware imperfections. To clarify the evaluation framework, we explicitly distinguish between two sets of parameters: the target gate parameters $(\theta_1,\theta_2,\varphi)$, which are strictly fixed by the desired quantum operation, and the hardware parameters $(L_1, T, \eta_1)$, which serve as the independent variables swept to map out the system's noise characteristics. The remaining parameter, the internal PSA gain $G_1$, acts as a dependent variable that adaptively scales to satisfy the required gain constraint dictated by Eq.~\eqref{required-gain}. 

To establish the analytical baseline for our evaluation, we first introduce a simplifying assumption regarding the input resource states. Specifically, to isolate the hardware-induced noise, we focus on the excess noise contributions by assuming an ideal EPR resource, where the first term in the brackets of Eq.~\eqref{output-GQT} vanishes. This assumption is physically justified because the excess noise induced by the feedforward circuit imperfections originates from vacuum fluctuations that are independent of the EPR resource state, so that the two contributions add independently at the level of the covariance matrix. Strictly speaking, at finite gain the EPR resource also enters the output through the de-amplified quadratures, yielding a small additional resource-dependent term that is not removed in the ideal-resource limit (see Appendix~\ref{Residual Noise Contributions from Non-Amplified Quadratures}); even for EPR resources generated from highly squeezed states of 25 and 30~dB, this term remains at most $-30$ and $-25$~dB relative to the vacuum level at the output for $T=0.99$, which is small compared with the hardware-induced noise evaluated below, and its impact is discussed in Sec.~\ref{sec:Discussion}. Therefore, excluding all resource-dependent contributions establishes a clean analytical baseline, allowing us to focus exclusively on the physical mechanisms and intuitive behaviors of the hardware-induced noise in the following discussion.

This hardware-induced noise behavior is fundamentally governed by two critical physical factors within the circuit block: the internal loss of the PSA and the transmittance $T$ of the displacement beam splitter. Specifically, while increasing $T$ closer to unity directly suppresses the feedforward-induced noise, it simultaneously escalates the required PSA gain. This elevated gain successfully counteracts the downstream propagation loss; however, it cannot overcome the internal loss of the PSA, which establishes an absolute, unavoidable noise floor due to the simultaneous amplification and attenuation occurring inside the non-ideal medium. Consequently, to approach the ideal lossless operation as modeled in Fig.~\ref{fig4}, the critical architectural requirement is to minimize the internal loss of the PSA while maximizing $T$ within the achievable gain range.

Appendices~\ref{Evolution of Covariance Matrices under Quantum Operations}, \ref{Physical Modeling of PSA with Internal Propagation Loss}, and \ref{app:noise-derivation} detail the covariance matrix evolution from the initial PSA stages through the balanced beam splitter and the final displacement coupling, yielding the excess noise covariance matrix $V_{\text{out}}$ at the output:
\begin{equation}\label{single-stage-total-noise}
    V_{\text{out}} = \frac{1-T}{2} \sum_{k=1}^{2} W_{\text{Rot}} \left[ (1-L_1) V_{\xi, \text{PSA}, k} + V_{\xi, \text{loss}} \right] W_{\text{Rot}}^T.
\end{equation}
Here, the explicit functional arguments are suppressed for notational brevity. The summation over $k=1,2$ represents the noise accumulated from the two parallel paths. The matrix $W_{\text{Rot}}$ denotes the phase rotation $W_{\text{Rot}}(\varphi)$ that aligns the correction field, $V_{\xi, \text{PSA}, k}$ is the intrinsic excess noise of the PSA in path $k$ (parameterized by the phase axis $\theta_k$, gain $G_1$, and internal efficiency $\eta_1$), and $V_{\xi, \text{loss}}$ represents the vacuum fluctuations injected via the consolidated external loss $L_1$. Because phase rotations commute with attenuation and losses following the 50:50 beam splitter can be mapped to equivalent losses at the inputs, all downstream propagation losses after the PSA are consolidated into the single loss parameter $L_1$ preceding the rotation stage.

\begin{figure*}[!ht] 
    \centering
    
    \begin{tikzpicture}
        \node[anchor=south west, inner sep=0] (image_a) at (0,0) {\includegraphics[width=0.6\textwidth]{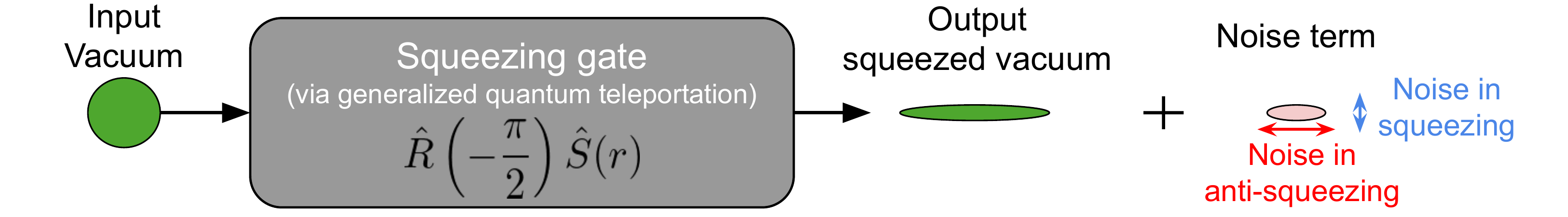}};
        \node[anchor=north west, xshift=-2em, yshift=0em] at (image_a.north west) {\textsf{(a)}};
    \end{tikzpicture}

    \vspace{1.5em} 

    \begin{minipage}[t]{0.48\textwidth}
        \vspace{0pt} 
        \centering
        \begin{tikzpicture}
            \node[anchor=south west, inner sep=0] (image_b) at (0,0) {\includegraphics[width=1\linewidth]{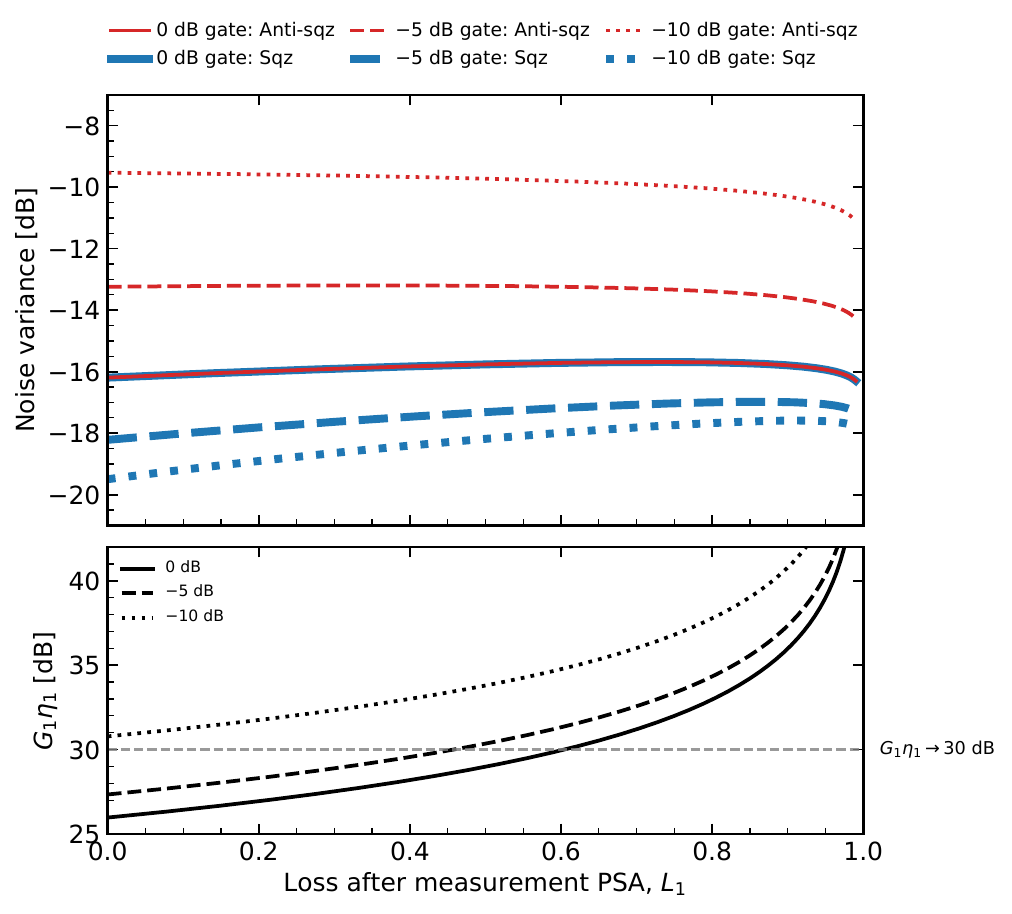}};
            \node[anchor=north west, xshift=0em, yshift=0em] at (image_b.north west) {\textsf{(b)}};
        \end{tikzpicture}

        \vspace{0.5em}

        \begin{tikzpicture}
            \node[anchor=south west, inner sep=0] (image_c) at (0,0) {\includegraphics[width=1\linewidth]{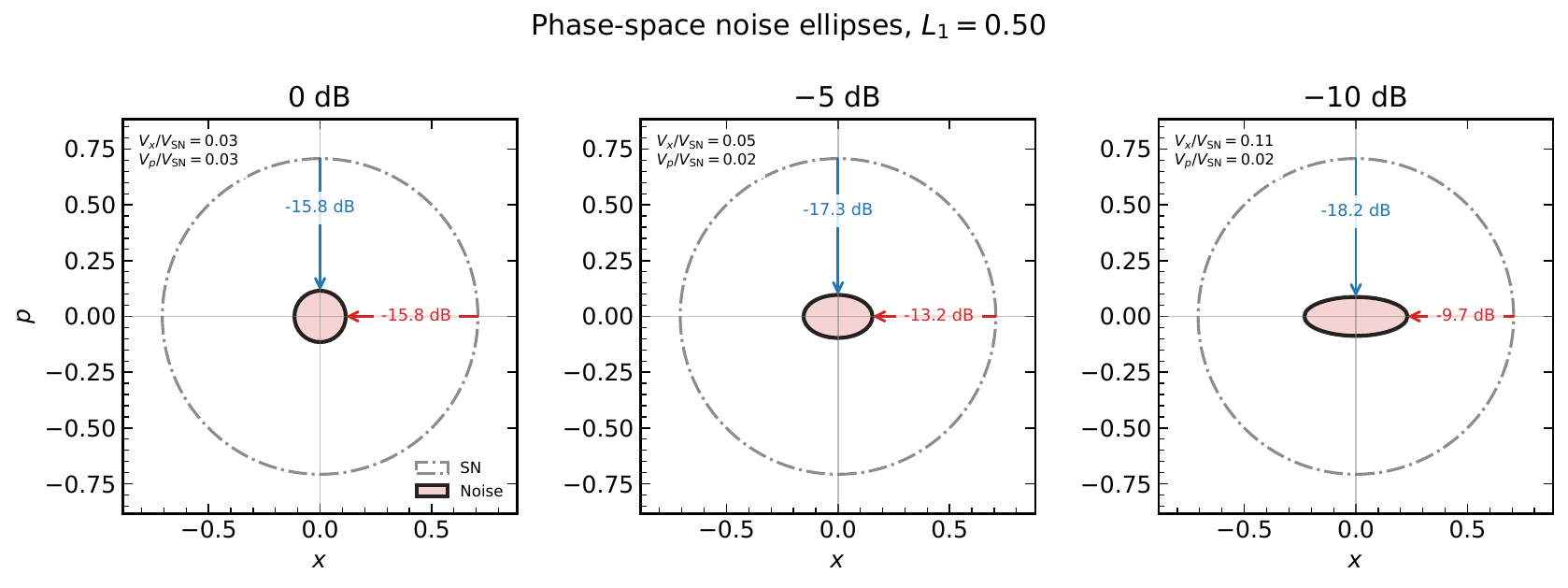}};
            \node[anchor=north west, xshift=0em, yshift=1em] at (image_c.north west) {\textsf{(c)}};
        \end{tikzpicture}
    \end{minipage}%
    \hfill 
    \begin{minipage}[t]{0.48\textwidth}
        \vspace{0pt} 
        \centering
        \begin{tikzpicture}
            \node[anchor=south west, inner sep=0] (image_d) at (0,0) {\includegraphics[width=1\linewidth]{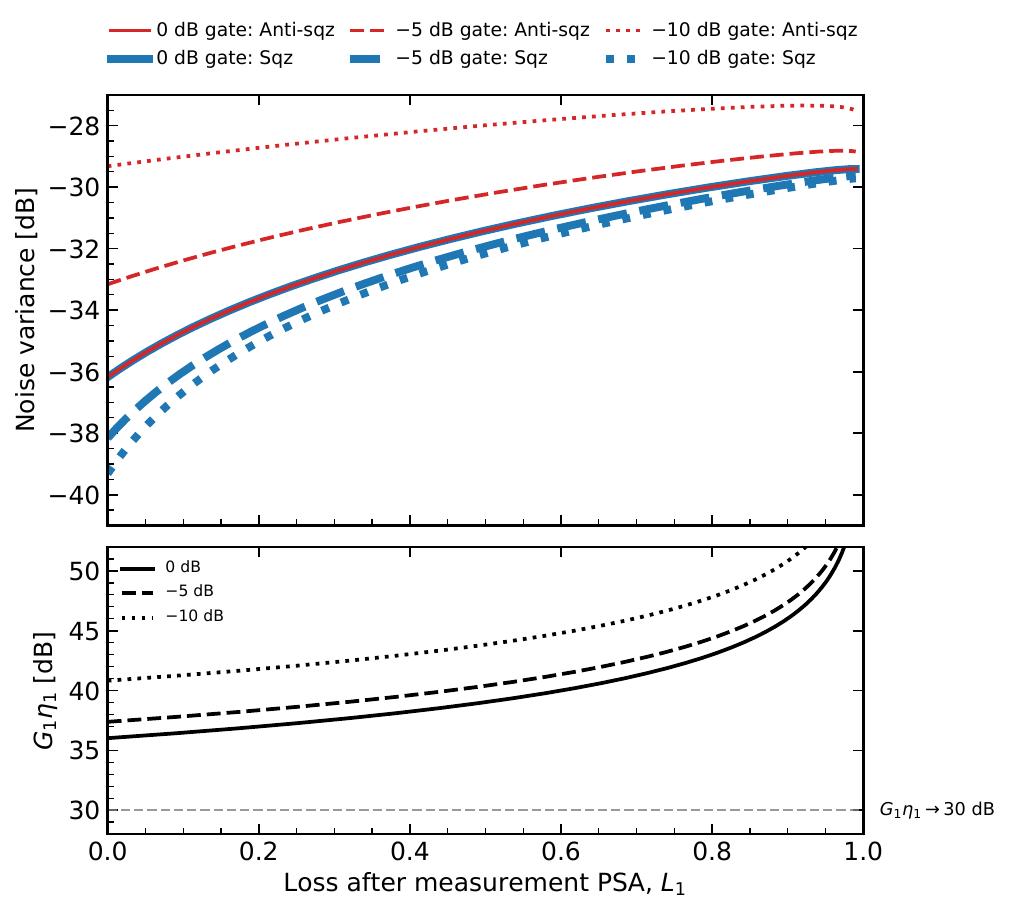}};
            \node[anchor=north west, xshift=0em, yshift=0em] at (image_d.north west) {\textsf{(d)}};
        \end{tikzpicture}

        \vspace{0.5em}

        \begin{tikzpicture}
            \node[anchor=south west, inner sep=0] (image_e) at (0,0) {\includegraphics[width=1\linewidth]{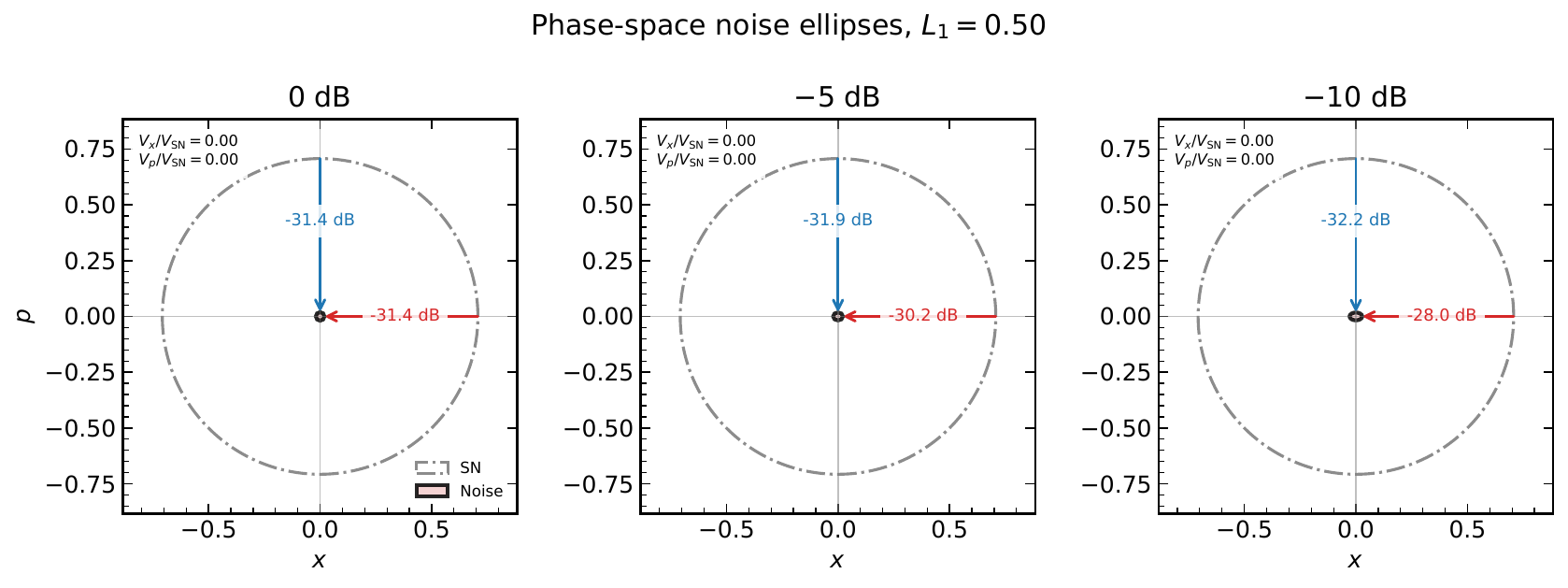}};
            \node[anchor=north west, xshift=0em, yshift=1em] at (image_e.north west) {\textsf{(e)}};
        \end{tikzpicture}
    \end{minipage}

    \vspace{0.8em} 
    
    \caption{Numerical analysis of the additional noise covariance $V_{\text{out}}$ [Eq.~\eqref{single-stage-total-noise}] for a single-stage building block. 
    (a) Conceptual output decomposition into the target state and accumulated noise. 
    (b)--(c) Performance under realistic parameters ($\eta_1 = 0.93, T = 0.99$): 
    (b) Additional noise power (top; noise variances evaluated independently of gain saturation) and required net PSA gain $G_1\eta_1 = G/(1-L_1)$ derived from Eq.~\eqref{required-gain} (bottom) versus external loss $L_1$ for $0$ (solid), $-5$ (dashed), and $-10$~dB (dotted) squeezing gates, where red and blue curves indicate noise in the anti-squeezing and squeezing axes, respectively. The horizontal gray dashed line marks the $30$~dB threshold, above which the target operation is physically difficult due to gain deficiency. 
    (c) Phase-space noise ellipses at $L_1=0.5$, showing geometric alignment locked along the $x$-axis relative to the shot-noise (SN) level, where the pink regions represent the amount of noise and the dashed lines indicate the SN region.
    (d)--(e) Performance under idealized conditions ($\eta_1 = 0.999, T = 0.999$): 
    (d) Additional noise variances (top; evaluated independently of gain saturation) and required net PSA gain $G_1\eta_1$ (bottom) versus external loss $L_1$. 
    (e) Phase-space noise ellipses at $L_1=0.5$, demonstrating near-complete suppression of the intrinsic noise bounds relative to the SN level, where the pink regions (amount of noise) are visually suppressed well below the SN region (dashed lines) as defined in (c).}
    \label{fig5}
\end{figure*}
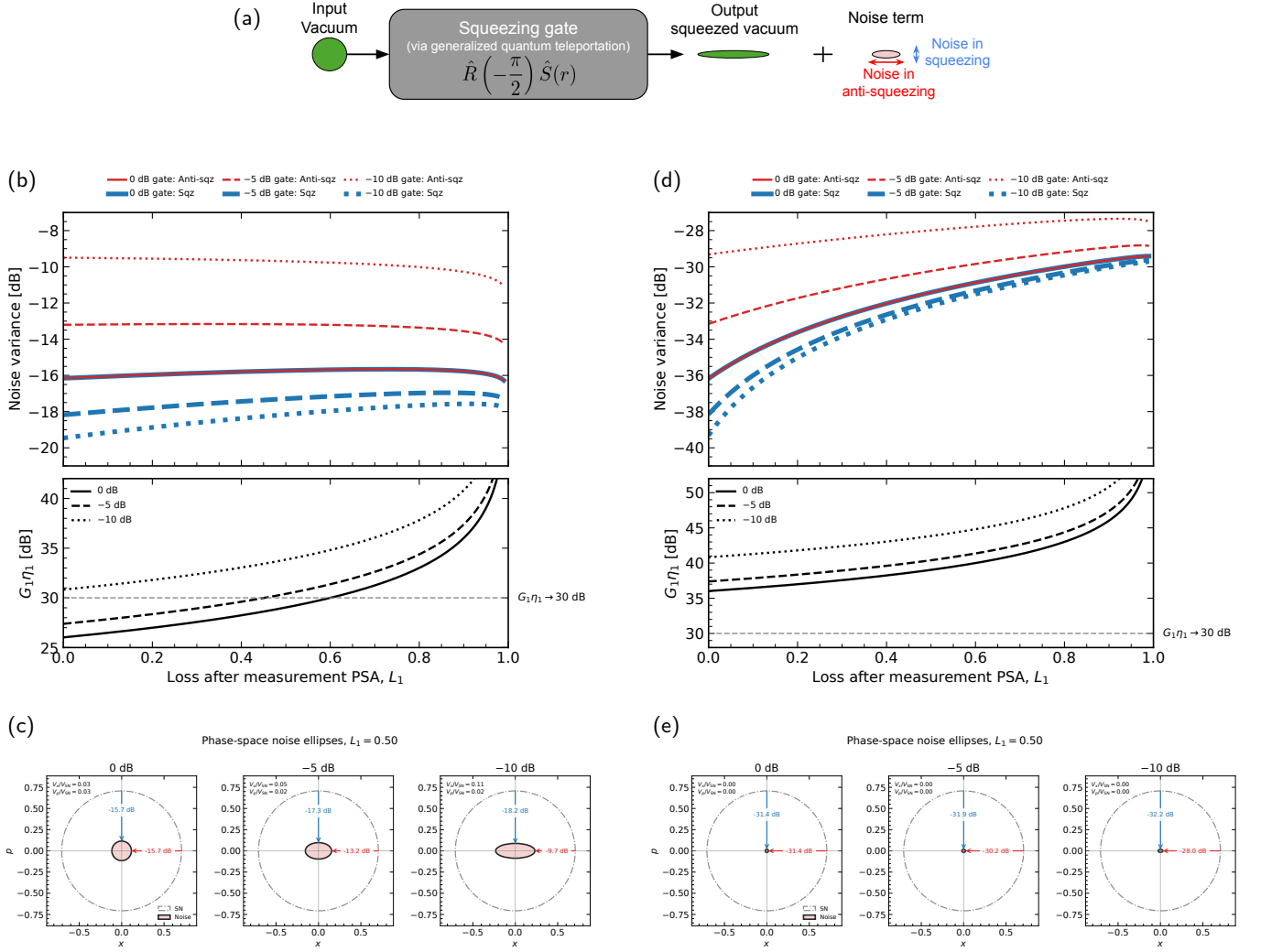

To evaluate Eq.~\eqref{single-stage-total-noise} in a physical context, we adopt a specific configuration to implement a squeezing operation that captures the geometric alignment and measurement-phase-dependent noise scaling of general adaptive operations without losing generality. The operation $\hat{R}(-\pi/2)\hat{S}(r)$ is realized with the measurement bases $\theta_1=\arctan(e^{-r})$ and $\theta_2=-\arctan(e^{-r})$, implementing anti-squeezing along the $x$-axis with $(\theta_1+\theta_2)/2 = 0$ for a vacuum input [Fig.~\ref{fig5}(a)]. Note that while this specific baseline assumes a positive squeezing parameter ($r > 0$), the non-positive regime ($r \le 0$) follows the identical physical scaling simply with an additional $\pi/2$ phase rotation. This setup preserves generality because any target operation decomposes into an anti-squeezing axis oriented at an angle of $(\theta_1+\theta_2)/2$ relative to the $x$-axis, along with a squeezing parameter determined by the relative measurement phase $(\theta_1-\theta_2)$ (see Appendix~\ref{Derivation of generalized quantum teleportation}). Crucially, as detailed in Appendix~\ref{app:noise-derivation}, the spatial orientation of the accumulated noise tensor is fundamentally governed by this average phase $(\theta_1+\theta_2)/2$—aligning precisely with the anti-squeezing axis—while its overall magnitude scales directly as a function of the relative measurement phase $(\theta_1 - \theta_2)$. Ultimately, the essence of preserving generality relies solely on two physical pillars: the geometric alignment between the anti-squeezing axis and the major axis of the noise tensor, and the deterministic dependency of both the squeezing parameter and the noise magnitude on the relative measurement phase. Therefore, the current configuration, which merely locks the primary noise and anti-squeezing directions along the $x$-axis by setting $(\theta_1+\theta_2)/2 = 0$ [Fig.~\ref{fig5}(a)], faithfully and comprehensively represents the noise behavior of general adaptive operations.

In quantifying the noise scaling, we treat the external propagation loss $L_1$ as the primary independent variable. To preserve the target operation, the total effective gain—determined by the combination $G=G_1\eta_1(1-L_1)$—must satisfy the condition dictated by $G$ in Eq.~\eqref{required-gain}. Thus, the adaptive PSA gain $G_1$ scales to compensate for the increased loss $L_1$.

We analyze how component quality bounds the total output noise by evaluating the fixed parameters $\eta_1$ and $T$ under two distinct regimes: current experimental capabilities (Fig.~\ref{fig5}(b) and (c)) and idealized physical limits (Fig.~\ref{fig5}(d) and (e)). For the realistic case (Fig.~\ref{fig5}(b) and (c)), the parameters are anchored to the experiment in \cite{suzuki2026ultrafastallopticalquantumteleportation}, which utilizes a 45-mm periodically poled lithium niobate (PPLN) waveguide. In this configuration, the internal efficiency is $\eta_1 = 0.93$ (limited by waveguide insertion losses \cite{10.1063/5.0063118, 10.1063/5.0144385}) and the displacement beam splitter transmittance is $T = 0.99$, with an achievable single-stage PSA gain cap of $G_1\eta_1 = 30$~dB. Throughout this paper, we refer to $G_1$ as the internal PSA gain, to $G_1\eta_1$ as the net PSA gain including the internal loss, and to $G$ in Eq.~\eqref{required-gain} as the total effective gain of the cascade [$G=G_1\eta_1(1-L_1)$ for the single-stage configuration]. Figure~\ref{fig5}(b) plots the excess noise power and the required net PSA gain $G_1\eta_1$ against the external propagation loss $L_1$, where the 30~dB threshold is indicated by a horizontal gray dashed line to denote the operational boundary of a single amplifier stage. Figure~\ref{fig5}(c) visualizes the noise distribution in phase space at $L_1 = 0.5$ (a typical loss for electro-optic phase modulators), illustrating the geometric dependence of the noise orientation on the cooperative phase settings. 

In addition to this realistic analysis, the ultimate physical limit of the architecture is subsequently investigated by evaluating its performance in a near-ideal regime where $\eta_1, T \to 0.999$ (Fig.~\ref{fig5}(d) and (e)). This idealized scenario assumes that extrinsic scattering losses are completely eliminated from the same 45-mm PPLN waveguide geometry, leaving only the intrinsic material attenuation of approximately 0.1~dB/m \cite{10.1063/5.0141436}.

Comparing Fig.~\ref{fig5}(b) and Fig.~\ref{fig5}(d) provides several physical insights. First, both configurations exhibit robustness against external propagation loss $L_1$; provided that a matching adaptive gain $G_1$ is supplied, the excess noise injected from downstream losses is suppressed. Second, in the idealized limit [Fig.~\ref{fig5}(d)], the intrinsic noise bounds imposed by $\eta_1$ and $T$ decrease to near-zero levels. This behavior validates that mitigating the internal propagation loss of the first-stage PSA ($\eta_1 \to 1$) and increasing the feedforward transmittance ($T \to 1$) are critical to approaching lossless operation. However, in this idealized limit or under large external loss conditions, the required net PSA gain scales dramatically, leading to gain saturation that cannot be circumvented by a single-stage optical amplifier. Overcoming this gain bottleneck establishes the physical necessity of the multi-stage amplification architecture proposed in the subsequent section.

\subsection{Noise Evaluation of the Multi-Stage Building Block}

\begin{figure*}[!ht]
    \centering
    \includegraphics[width=0.8\linewidth]{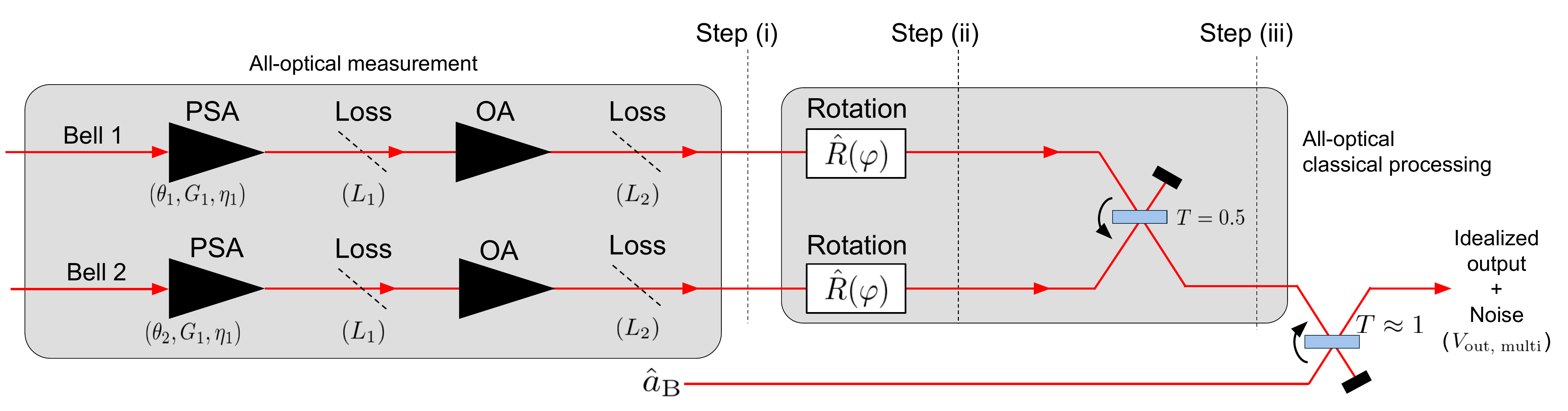}
    \caption{Mathematical modeling of the multi-stage building block with gain compensation incorporating experimental imperfections. The setup parameterizes the primary and secondary optical amplifiers (labeled as OA; gains $G_1, G_2$, internal efficiencies $\eta_1, \eta_2$, phase $\theta_k$), subsequent propagation losses $L_1, L_2$, and feedforward phase rotation $\varphi$. It further includes a 50:50 BS and a high-transmittance BS ($T \sim 1$) for displacement coupling with mode $\hat{a}_B$, yielding the final output noise covariance matrix $V_{\text{out,multi}}^{\text{OA}}$ [Eq.~\eqref{multi-stage-psa-noise}, \eqref{multi-stage-pia-noise}].}
    \label{fig6}
\end{figure*}

To overcome the single-stage gain limitation, a multi-stage configuration is evaluated to distribute the required gain (Fig.~\ref{fig6}). Maintaining the same evaluation framework, the gate parameters $(\theta_k, \varphi)$ remain strictly fixed, while the multi-stage hardware parameters $(L_1, L_2, \eta_1, \eta_2)$ serve as the independent variables representing circuit imperfections. Under this setup, the individual stage gains function as dependent variables coordinated to satisfy the total effective gain constraint dictated by Eq.~\eqref{required-gain}. Specifically, the total combination $G_1\eta_1(1-L_1)G_2\eta_2(1-L_2)$ (where $\eta_2=1$ for an ideal phase-insensitive stage) is adaptively controlled, meaning that the secondary gain $G_2$ automatically scales according to the allocated first-stage gain $G_1$ and the hardware parameters. This secondary amplification stage can be implemented via either a PSA or a phase-insensitive amplifier (PIA).

\subsubsection{Multi-Stage Configuration with Dual PSA}

To eliminate isotropic excess noise, a dual-PSA configuration is analyzed where both stages utilize phase-sensitive amplification. By denoting the intermediate noise covariance matrix accumulated after the first-stage propagation loss $L_1$ as
\begin{equation}\label{inter-stage-noise}
    V_{\text{out}, k} = (1-L_1) V_{\xi, \text{PSA}, k} + V_{\xi, \text{loss1}},
\end{equation}
and noting that the secondary PSA selectively amplifies the target quadrature without injecting spontaneous emission noise, the total multi-stage output noise covariance matrix $V_{\text{out, multi}}^{\text{PSA}}$ is formulated as:
\begin{widetext}
\begin{equation}\label{multi-stage-psa-noise}
    V_{\text{out, multi}}^{\text{PSA}} = \frac{1-T}{2} \sum_{k=1}^{2} W_{\text{Rot}} \Bigl[ (1-L_2) \Bigl\{ W_{\text{PSA2}, k} V_{\text{out}, k} W_{\text{PSA2}, k}^T + V_{\xi, \text{PSA2}} \Bigr\} + V_{\xi, \text{loss2}} \Bigr] W_{\text{Rot}}^T,
\end{equation}
\end{widetext}
where the explicit parameter dependencies are again suppressed for notational simplicity. Here, $W_{\text{Rot}}$ and $V_{\text{out}, k}$ are defined as in the single-stage baseline, while $V_{\xi, \text{loss1}}$ and $V_{\xi, \text{loss2}}$ represent the vacuum fluctuations injected via the propagation losses $L_1$ and $L_2$, respectively. The matrices $W_{\text{PSA2}, k}$ and $V_{\xi, \text{PSA2}}$ denote the transformation $W_{\text{PSA2}}(\theta_k, G_2, \eta_2)$ and the internal imperfections of the secondary PSA in path $k$, whose amplification axis is adaptively aligned to $\theta_k$.

The numerical results in Fig.~\ref{fig7}(a) and (b) are obtained using the parameters from \cite{suzuki2026ultrafastallopticalquantumteleportation} with the inter-stage loss fixed at $L_1 = 0.1$, which is determined by the coupling loss to the latter PSA, and the second-stage PSA is assumed to have the same internal efficiency as the first stage, $\eta_2=\eta_1=0.93$. The gains $G_1$ and $G_2$ are adaptively balanced; for $G_1\eta_1 < 30$~dB, $G_1$ counteracts the loss while $G_2 = 1$. When the total required gain exceeds 30~dB, $G_1\eta_1$ is clamped at 30~dB, and the remaining gain burden is shifted to $G_2$. 

As shown in Fig.~\ref{fig7}(a) and (b), the dual-PSA scheme maintains a low noise profile comparable to the single-stage baseline. By bypassing the fundamental noise penalty of conventional linear amplifiers, this noiseless secondary amplification suppresses the downstream vacuum fluctuations $V_{\xi, \text{loss}}(L_2)$, preserving the quantum state characteristics under propagation losses.

\subsubsection{Multi-Stage Configuration with a PIA}

Alternatively, a PIA-assisted configuration is analyzed. As detailed in Appendix~\ref{app:noise-derivation}, while a PIA provides isotropic gain without requiring pump-phase alignment, it introduces a quantum noise penalty. Using the intermediate noise matrix $V_{\text{out}, k}$ defined in Eq.~\eqref{inter-stage-noise}, the total output noise covariance matrix $V_{\text{out, multi}}^{\text{PIA}}$ is written as:
\begin{widetext}
\begin{equation}\label{multi-stage-pia-noise}
    V_{\text{out, multi}}^{\text{PIA}} = \frac{1-T}{2} \sum_{k=1}^{2} W_{\text{Rot}} \Bigl[ (1-L_2) \Bigl\{ W_{\text{PIA}} V_{\text{out},k} W_{\text{PIA}}^T + V_{\xi, \text{PIA}} \Bigr\} + V_{\xi, \text{loss2}} \Bigr] W_{\text{Rot}}^T,
\end{equation}
\end{widetext}
where the explicit parameter dependencies are again suppressed for consistency. Here, $W_{\text{Rot}}$, $V_{\text{out}, k}$, and $V_{\xi, \text{loss2}}$ are defined as in the previous sections. The operators $W_{\text{PIA}}$ and $V_{\xi, \text{PIA}}$ represent the isotropic transformation $W_{\text{PIA}}(G_2)$ and the intrinsic quantum noise penalty of the PIA, respectively. This added noise is assumed to enforce a minimum noise figure that asymptotically approaches the standard quantum limit (SQL) of 3~dB in the high-gain limit ($G_2 \to \infty$).

\begin{figure*}[!ht]
    \centering
    
    \begin{minipage}[t]{0.48\textwidth}
        \vspace{0pt} 
        \centering
        
        \begin{tikzpicture}
            \node[anchor=south west, inner sep=0] (image_a) at (0,0) {\includegraphics[width=1\linewidth]{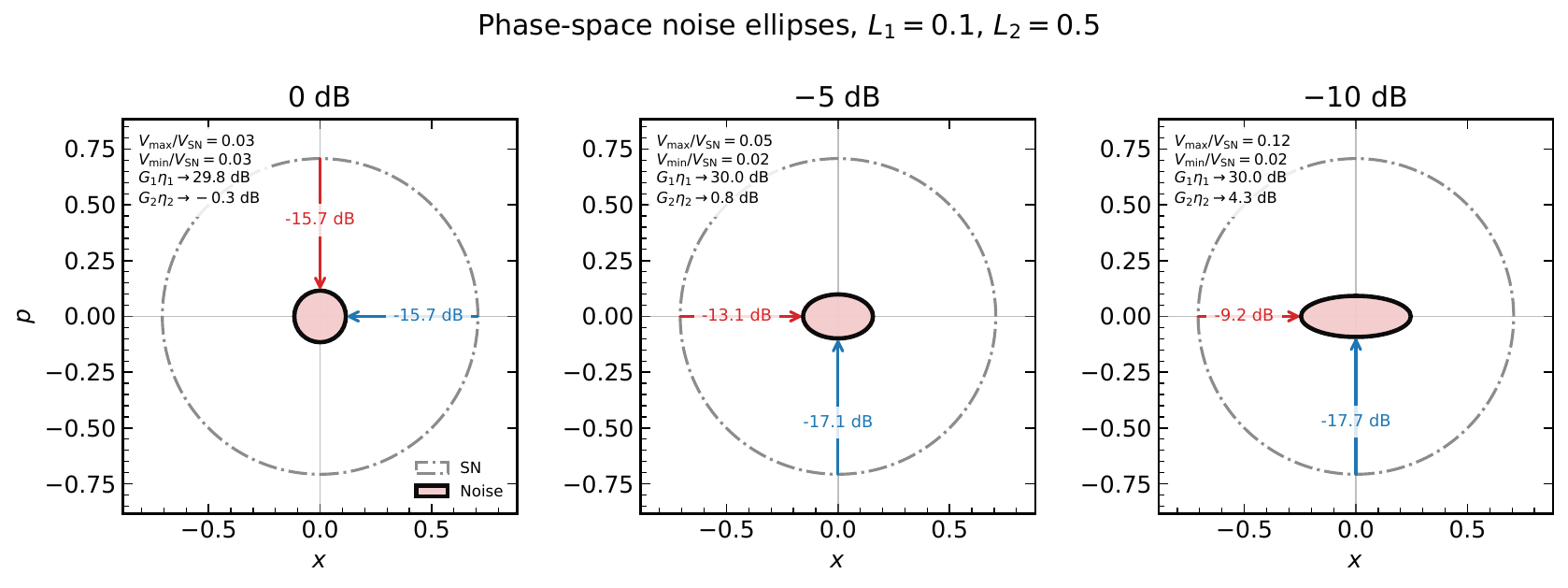}};
            \node[anchor=north west, xshift=0em, yshift=0em] at (image_a.north west) {\textsf{(a)}};
        \end{tikzpicture}

        \vspace{0.5em}

        \begin{tikzpicture}
            \node[anchor=south west, inner sep=0] (image_b) at (0,0) {\includegraphics[width=1\linewidth]{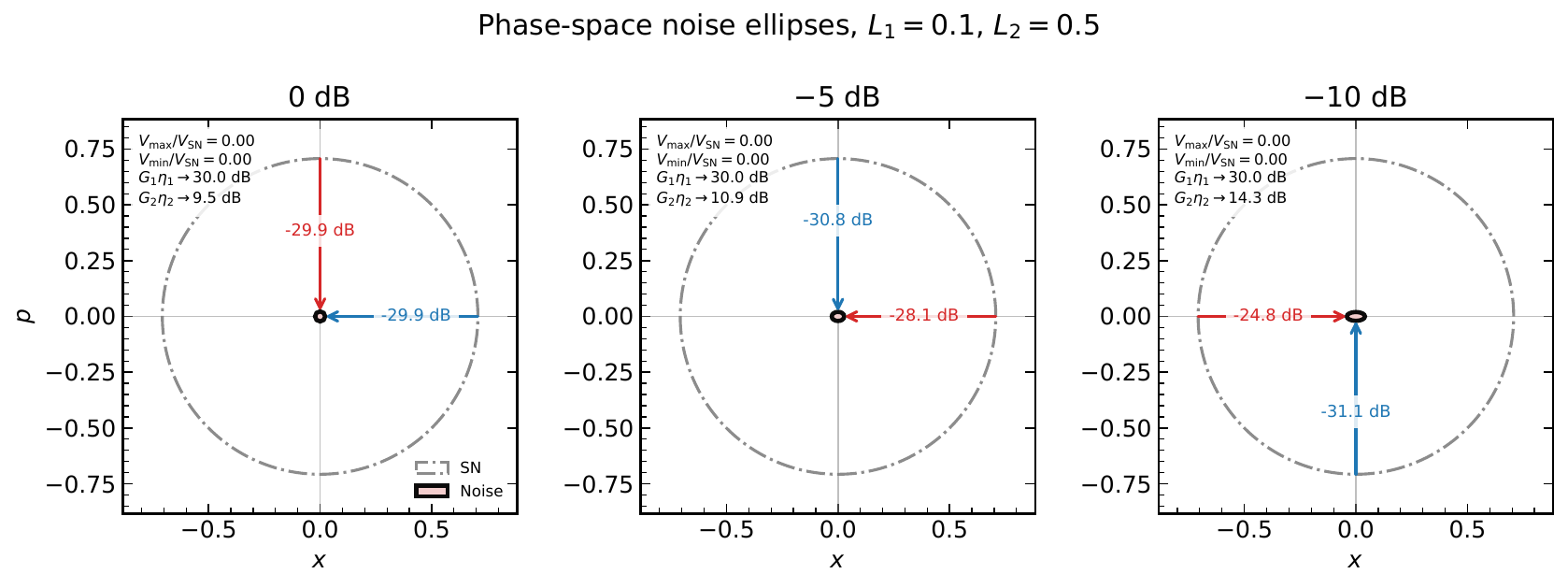}};
            \node[anchor=north west, xshift=0em, yshift=0em] at (image_b.north west) {\textsf{(b)}};
        \end{tikzpicture}
    \end{minipage}%
    \hfill 
    \begin{minipage}[t]{0.48\textwidth}
        \vspace{0pt} 
        \centering
        
        \begin{tikzpicture}
            \node[anchor=south west, inner sep=0] (image_c) at (0,0) {\includegraphics[width=0.95\linewidth]{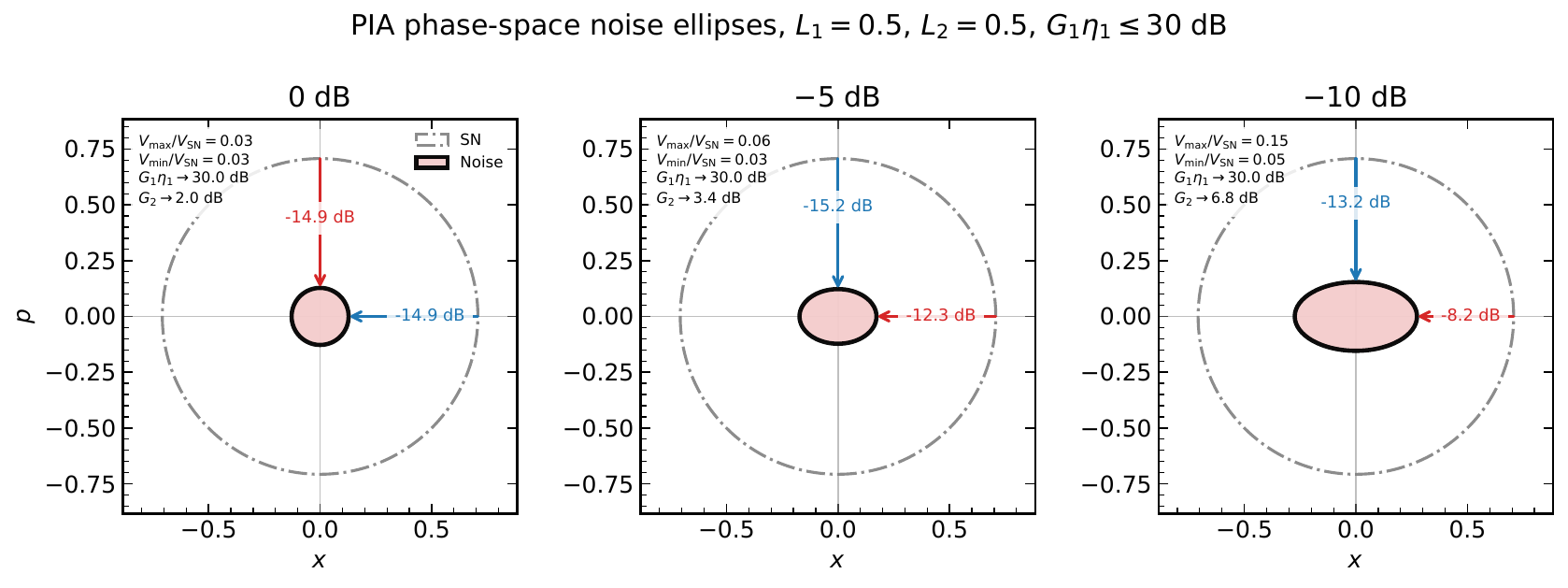}};
            \node[anchor=north west, xshift=0em, yshift=0em] at (image_c.north west) {\textsf{(c)}};
        \end{tikzpicture}

        \vspace{0.5em}

        \begin{tikzpicture}
            \node[anchor=south west, inner sep=0] (image_d) at (0,0) {\includegraphics[width=0.95\linewidth]{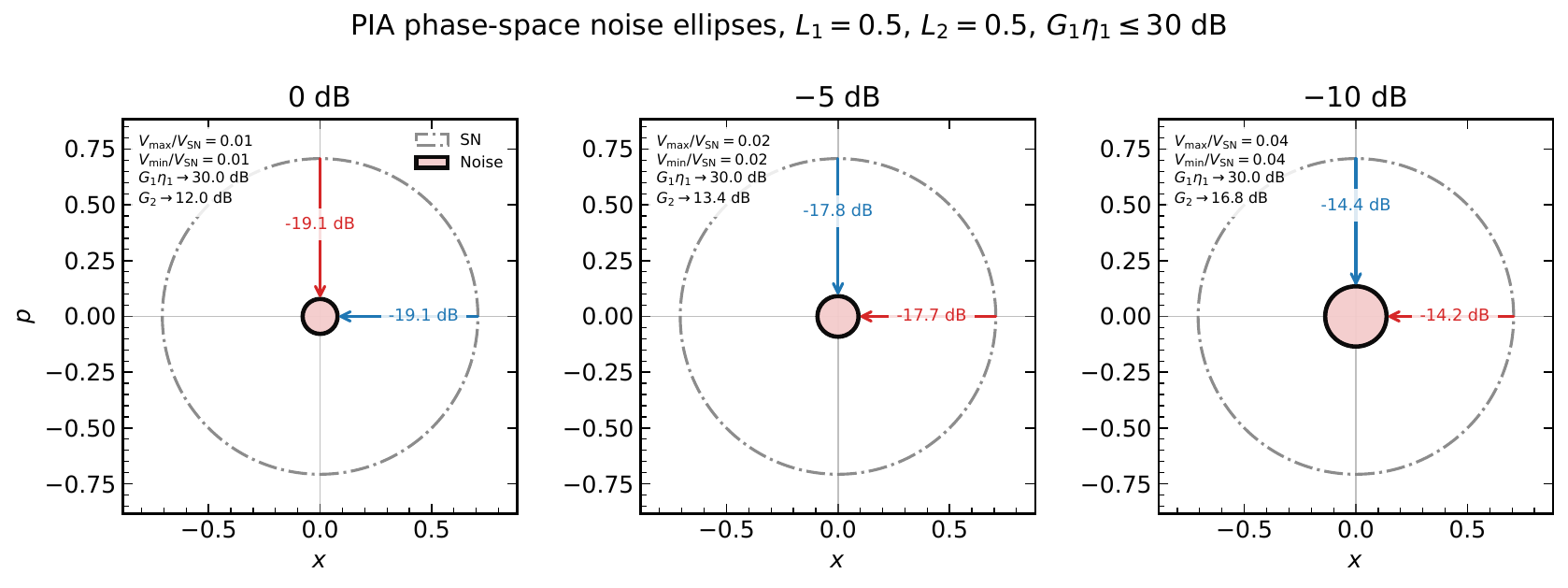}};
            \node[anchor=north west, xshift=0em, yshift=0em] at (image_d.north west) {\textsf{(d)}};
        \end{tikzpicture}
    \end{minipage}

    \vspace{0.8em} 
    
    \caption{Noise performance and loss dependence of the multi-stage configurations. 
    (a)--(b) Dual-PSA configuration (inter-stage loss $L_1=0.1$) under (a) realistic parameters ($\eta_1 = \eta_2 = 0.93, T = 0.99$) and (b) idealized components ($\eta_1 = \eta_2 = T = 0.999$). 
    (c)--(d) PIA configuration (inter-stage loss $L_1=0.5$) under (c) realistic parameters ($\eta_1 = 0.93, T = 0.99$) and (d) idealized components ($\eta_1 = 0.999, T = 0.999$).
    Note that the two configurations differ not only in the amplifier physics but also in the assumed inter-stage loss $L_1$, which reflects different candidate layouts; the numerical difference between (a)--(b) and (c)--(d) should therefore not be attributed solely to the amplifier type. The PIA is modeled as quantum limited (see Appendix~\ref{Evolution of Covariance Matrices under Quantum Operations}).}
    \label{fig7}
\end{figure*}
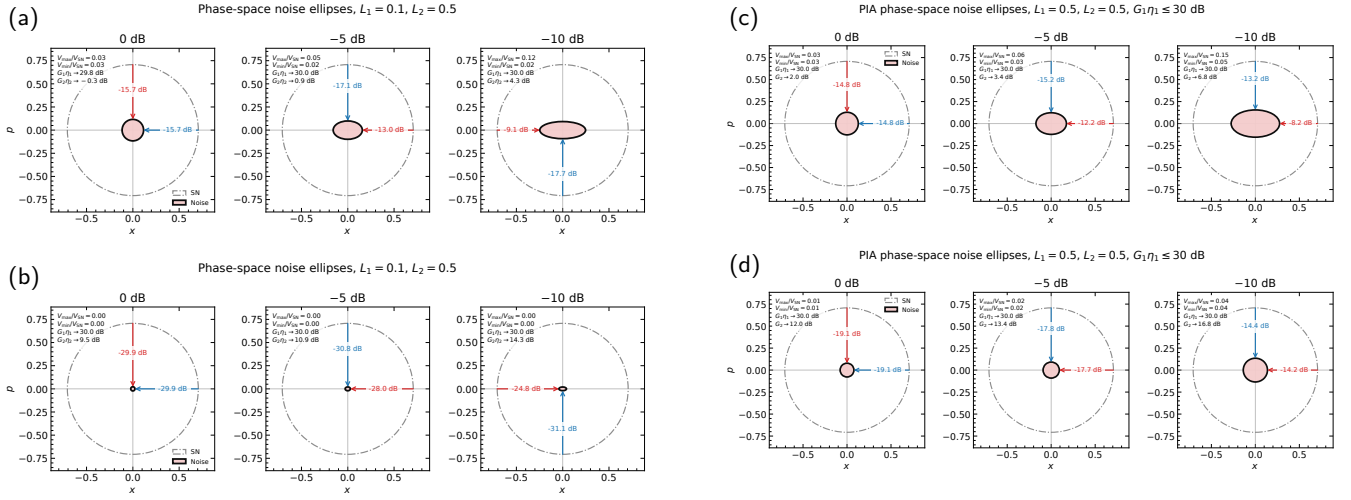

The numerical analysis in Fig.~\ref{fig7}(c) and (d) assumes $L_1 = 0.5$, a coupling loss achievable in integrated optical platforms equipped with on-chip PIAs \cite{app9081588,Sobhanan:22}. Symmetrically to the dual-PSA scheme, the gains are balanced by clamping $G_1\eta_1$ at 30~dB and shifting the remaining burden to $G_2$.

While the noise scaling behavior remains similar to the single-stage architecture regarding the suppression of downstream loss $L_2$, the PIA configuration introduces a prominent intrinsic noise penalty. When the required secondary gain $G_2$ escalates—driven either by downstream losses or the demands of high-degree squeezing gates—this quantum noise penalty accumulates within the device. Consequently, while this additional noise floor is acceptable for moderate operations, it degrades gate performance as the target squeezing approaches deeper regimes, such as 10~dB.

\section{Discussion}
\label{sec:Discussion}

We now assess the feasibility of integrating the proposed AOFF architectures into practical quantum computing frameworks by examining the compatibility of the calculated noise floors with error-correction requirements, the technical viability of the hardware parameters, and the prospects for ultrafast processing.

\subsection{Noise Resilience and Hardware Parameter Validation}

To evaluate architectural viability independently of specific downstream quantum tasks, the accumulated output noise is mapped back to the input port to quantify the preservation of quantum information. Using the covariance matrix formalism, the total output covariance $V_{\text{out}}$ is refactored as:
\begin{equation}\label{IO-covariance}
\begin{aligned}
    V &= W V_{\text{in}} W^T + V_{\text{ex}} \\
    &= W \left( V_{\text{in}} + W^{-1} V_{\text{ex}} (W^T)^{-1} \right) W^T \\
    &= W \left( V_{\text{in}} + V_{\text{in, eq}} \right) W^T
\end{aligned}
\end{equation}
where $W$ is the target ideal transformation—comprising the squeezing operator $\hat{S}(r)$ and a phase rotation—$V_{\text{ex}}$ is the hardware-induced excess noise at the output, and $V_{\text{in, eq}} \equiv W^{-1} V_{\text{ex}} (W^T)^{-1}$ defines the equivalent input noise (EIN) covariance matrix.

This refactoring shows that the noisy feedforward process is mathematically equivalent to injecting an equivalent noise covariance $V_{\text{in, eq}}$ directly at the input, followed by an ideal, lossless operation $W$. Because $W^{-1}$ executes the inverse transformation, the output excess noise components along the squeezed ($V_{\text{sq}}$) and anti-squeezed ($V_{\text{anti}}$) quadratures scale back to the input as $e^{2r} V_{\text{sq}}$ and $e^{-2r} V_{\text{anti}}$, respectively. The maximum EIN, defined as $\max\{V_{\text{sq}} e^{2r}, V_{\text{anti}} e^{-2r}\}$, provides a task-independent metric for input state preservation. As summarized across various target squeezing levels in Table~\ref{tab:all_architecture_comparison_summary}, this maximum EIN increases monotonically with gate demand, illustrating the trade-offs inherent to each architectural configuration.

\begin{table*}[!ht]
\centering
\renewcommand{\arraystretch}{1.25}
\caption{Performance summary of target squeezing $S$, maximum equivalent input noise (EIN), and required gains. The Max EIN column displays the input-referred noise power defined as $\max\{V_{\text{sq}} e^{2r}, V_{\text{anti}} e^{-2r}\}$, with its left-hand symbols ($\checkmark$, $\times$) denoting noise suppression thresholds of $< -12$~dB, and $\ge -12$~dB, respectively. The gain column lists operational gains ($G_1\eta_1, G_2\eta_2$), with its left-hand symbols ($\checkmark$, $\times$) indicating whether each gain remains within or exceeds the $30$~dB empirical threshold at the first PSA. For the PSA-PIA scheme, the column displays the pure internal gain $G_2$ since $\eta_2$ is physically undefined. For the realistic dual-PSA configuration at $0$~dB, the required gain is provided by the first stage alone, so that the second PSA is not pumped ($G_2=0$~dB) and acts only as an insertion loss, which is indicated by a dash. The single-stage values are evaluated at $L_1=0.5$, and the dual-PSA and PSA-PIA schemes assume different inter-stage losses ($L_1=0.1$ and $0.5$, respectively; see Fig.~\ref{fig7}), with $L_2=0.5$; the PIA is modeled as quantum limited.}
\label{tab:all_architecture_comparison_summary}
\begin{tabular}{l c cc cc}
\toprule 
\textbf{Architecture} & \textbf{Target $S$ (dB)} & \multicolumn{2}{c}{\textbf{Max EIN (dB)}} & \multicolumn{2}{c}{\textbf{Gain of Amp. ($G_1\eta_1, G_2\eta_2$) (dB)}} \\ \midrule 
\textbf{Single-Stage} (Figs.~\ref{fig5}) & \multicolumn{5}{c}{\textit{Realistic Configuration ($\eta_1=0.93, T=0.99$)}} \\ \cmidrule(lr){2-6} 
                      & 0.00                     & $\checkmark$ & -15.8     & $\checkmark$ & $(29.0, *)$            \\
                      & 5.00                     & $\checkmark$ & -12.3     & $\times$     & $(30.4, *)$           \\
                      & 10.0                     & $\times$     & -8.20     & $\times$     & $(33.8, *)$           \\ \addlinespace
                      & \multicolumn{5}{c}{\textit{Idealized Configuration ($\eta_1=T=0.999$)}} \\ \cmidrule(lr){2-6}
                      & 0.00                     & $\checkmark$ & -31.4     & $\times$     & $(39.0, *)$           \\
                      & 5.00                     & $\checkmark$ & -26.9     & $\times$     & $(40.4, *)$           \\
                      & 10.0                     & $\checkmark$ & -22.2     & $\times$     & $(43.8, *)$           \\ \midrule
\textbf{Dual-PSA} (Fig.~\ref{fig7}) & \multicolumn{5}{c}{\textit{Realistic Configuration ($\eta_1=\eta_2=0.93, T=0.99$)}} \\ \cmidrule(lr){2-6}
                      & 0.00                     & $\checkmark$ & -15.7     & $\checkmark$ & $(29.8, \text{---})$            \\
                      & 5.00                     & $\checkmark$ & -12.1     & $\checkmark$ & $(30.0, 0.81)$            \\
                      & 10.0                     & $\times$     & -7.70     & $\checkmark$ & $(30.0, 4.25)$            \\ \addlinespace
                      & \multicolumn{5}{c}{\textit{Idealized Configuration ($\eta_1=\eta_2=T=0.999$)}} \\ \cmidrule(lr){2-6}
                      & 0.00                     & $\checkmark$ & -29.9     & $\checkmark$ & $(30.0, 9.48)$            \\
                      & 5.00                     & $\checkmark$ & -25.8     & $\checkmark$ & $(30.0, 10.85)$            \\
                      & 10.0                     & $\checkmark$ & -21.1     & $\checkmark$ & $(30.0, 14.29)$            \\ \midrule
\textbf{PSA-PIA} (Fig.~\ref{fig7})  & \multicolumn{5}{c}{\textit{Realistic Configuration ($\eta_1=0.93, T=0.99$)}} \\ \cmidrule(lr){2-6}
                      & 0.00                     & $\checkmark$ & -14.9     & $\checkmark$ & $(30.0, 2.00)$            \\
                      & 5.00                     & $\times$     & -10.2     & $\checkmark$ & $(30.0, 3.36)$            \\
                      & 10.0                     & $\times$     & -3.20     & $\checkmark$ & $(30.0, 6.80)$            \\ \addlinespace
                      & \multicolumn{5}{c}{\textit{Idealized Configuration ($\eta_1=T=0.999$)}} \\ \cmidrule(lr){2-6}
                      & 0.00                     & $\checkmark$ & -19.1     & $\checkmark$ & $(30.0, 12.04)$            \\
                      & 5.00                     & $\checkmark$ & -12.8     & $\checkmark$ & $(30.0, 13.40)$            \\
                      & 10.0                     & $\times$     & -4.40     & $\checkmark$ & $(30.0, 16.84)$            \\ \bottomrule 
\end{tabular}
\end{table*}

\subsubsection{Implications for Fault-Tolerant GKP Quantum Computation}

We evaluate the applicability of the proposed architecture within the context of GKP-based FTQC. As shown in Table~\ref{tab:all_architecture_comparison_summary}, under realistic hardware parameters, the PSA-based configurations (single-stage and dual-PSA) maintain the maximum EIN below $-12$~dB for target squeezing operations up to 5~dB ($-12.3$~dB and $-12.1$~dB at 5~dB, respectively, where the single-stage configuration requires a net PSA gain slightly above 30~dB). In contrast, the PSA-PIA configuration exceeds this level already at 5~dB ($-10.2$~dB), and all realistic configurations exceed it at 10~dB. Within this range of the PSA-based configurations, assuming an initial GKP state with an effective squeezing level of 12~dB—a resource specification aligned with recent experimental benchmarks \cite{ha2026generation12dbsqueezed}—the effective squeezing of the GKP qubit is preserved at approximately 9.0--9.1~dB post-feedforward. This remains above the 8.1~dB threshold estimated for the GKP surface code under realistic optical conditions \cite{PhysRevA.107.052414}, suggesting that the architecture is potentially compatible with GKP-based FTQC.

The above evaluation is based on the additive hardware-induced noise summarized in Table~\ref{tab:all_architecture_comparison_summary}. In addition, the exact input--output relation derived in Appendix~\ref{Residual Noise Contributions from Non-Amplified Quadratures} shows that the finite gain and the finite transmittance $T$ modify the feedforward output in two ways: the target operation $\mathbf{V}(\theta_1,\theta_2)$ is replaced by $\sqrt{T}\lambda\,\mathbf{V}(\theta_1',\theta_2')$ [Eq.~\eqref{eq:C_gate}], and an input-independent term originating from the EPR resource appears [Eq.~\eqref{eq:C_result}]. The former, referred back to the input, rescales the input state, and hence the GKP grid, by factors $\kappa_\pm=\sqrt{T}\lambda\,e^{\pm(r'-r)}$ along the principal axes, which satisfy $|1-\kappa_\pm|\le5\times10^{-3}$ for $T=0.99$ and $|1-\kappa_\pm|\le5\times10^{-4}$ for $T=0.999$ over the target squeezing range of $0$--$10$~dB. Such a deterministic rescaling of the grid, as also caused by optical losses, cannot simply be regarded as Gaussian random displacement noise~\cite{Harris2025LossGKP}, and its rigorous treatment within a fault-tolerance analysis is left for future work. Here, we estimate its impact by a simple conversion. For an approximate GKP state with peak variance $\Delta^2/2$ and a Gaussian envelope of variance $1/(2\Delta^2)$, the rescaling shifts a peak located at quadrature value $q$ by $(\kappa_\pm-1)q$; averaged over the envelope, this can be regarded as an additional peak broadening with variance $(1-\kappa_\pm)^2/(2\Delta^2)$. For the 12-dB GKP state ($\Delta^2\approx0.063$) and $T=0.99$, this additional variance is at most $-34$~dB relative to the vacuum level, degrading the effective squeezing by less than $0.03$~dB; for $T=0.999$, the degradation is below $0.001$~dB.

The latter, the resource-induced term, grows with the resource squeezing; we therefore evaluate it for EPR resources generated from highly squeezed states with squeezing levels of 25 and 30~dB to obtain a pessimistic estimate (see Appendix~\ref{Residual Noise Contributions from Non-Amplified Quadratures} and Table~\ref{tab:C}). For $T=0.99$, it increases the maximum EIN at the 5-dB target squeezing of the single-stage and dual-PSA configurations from $-12.3$ and $-12.1$~dB to $-12.2$ and $-12.0$~dB for a resource squeezing level of 25~dB and to $-11.9$ and $-11.8$~dB for 30~dB, respectively. Even in the latter case, the effective squeezing of the 12-dB GKP state after the feedforward remains at approximately $8.9$~dB or higher, which is still above the 8.1~dB threshold. Since the EIN increases monotonically with the target squeezing, this conclusion holds for all target operations up to 5~dB, including the Clifford operations discussed below. For $T=0.999$, the increase of the maximum EIN is below $0.2$~dB even for a resource squeezing level of 30~dB. These corrections arising from the finite gain and the finite transmittance therefore do not affect the above conclusion.

Moreover, the evaluated 5~dB dynamic range covers the core Gaussian operations required for GKP quantum computation. Because the Controlled-NOT gate decomposes into a Controlled-Z (CZ) gate and Hadamard gates, the necessary Clifford operations reduce to the Phase, Hadamard, and CZ gates. In the TDM-MBQC framework, a single-step quantum operation within the quad-rail lattice (QRL)~\cite{PhysRevA.93.062326,yokoyama2026fullstackanalogopticalquantum} is fundamentally described by a combination of static multimode interference via fixed beam splitters and localized single-mode GQTs, the primary focus of this study. By utilizing this single-step operational unit, these target operations are systematically implemented. For instance, each GQT stage executes a specific single-mode operation depending on its local measurement basis: it performs the Phase gate ($\exp(i\hat{x}^2/2)$) when configured to $\hat{V}(\pi/2,\arctan(1/2))$ defined in Eq.~\eqref{GQT-operation}, and the Hadamard gate ($\exp(i\pi\hat{a}^\dagger\hat{a}/2)$) when configured to $\hat{V}(3\pi/4,\pi/4)$~\cite{yokoyama2026fullstackanalogopticalquantum}. Meanwhile, the two-mode CZ gate ($\exp(i\hat{x}_1\hat{x}_2)$) is naturally realized by leveraging the fixed BS entangling network alongside localized shear operations, which are implemented by configuring the GQT measurement bases to $\hat{V}(\pi/2,\arctan(\pm1/2))$ across the respective modes. Finally, evaluating Eq.~\eqref{GQT-operation} demonstrates that to execute the Clifford operations required for GKP qubits within this TDM-MBQC framework, the gate operations required within the localized GQT stages demand a maximum internal squeezing magnitude of exactly $4.18$~dB (specifically for the CZ and Phase gate steps). Since the remaining constituent steps require no squeezing ($0$~dB), these results firmly validate that the entire required operational suite fits well within the evaluated 5~dB dynamic range.

Accordingly, with the PSA-based configurations, the proposed architecture supports the complete set of Clifford operations required for GKP computation without introducing a fidelity bottleneck under realistic losses. In the idealized limit, the hardware-induced noise of the feedforward process becomes negligible. These results establish the proposed architecture as a potential hardware platform for scaling GKP-based FTQC. To further solidify this framework, several important challenges remain to be addressed in future work. First, the architectural implementation of decoding circuits that incorporate nonlinear modulo measurements~\cite{PhysRevA.64.012310, PhysRevA.105.042427, Conrad2022gottesmankitaev}, alongside more advanced evaluations of the fault-tolerant threshold, is left for future investigations. Within this context, structural analyses of decoding circuits specifically tailored for high-level encoding architectures~\cite{Battistel2023RealTimeDecoding, Caune2026} remain an open and critical task. Furthermore, a comprehensive evaluation that fully incorporates the practical effects of non-zero noise in the EPR resources is also left as another crucial challenge.

\subsubsection{Physical Grounds of Component Parameters and Candidate Hardware Platforms}

To evaluate the experimental feasibility of the configurations summarized in Table~\ref{tab:all_architecture_comparison_summary}, we justify the physical basis of the adopted hardware parameters by benchmarking them against state-of-the-art platforms. The realistic baseline---characterized by a net PSA gain of $G_1\eta_1=30$~dB and an internal propagation efficiency of $\eta_1=0.93$---is directly anchored to recent experimental implementations of continuous-variable all-optical quantum teleportation \cite{suzuki2026ultrafastallopticalquantumteleportation,10.1063/5.0063118,10.1063/5.0144385}. In contrast, the idealized limit ($\eta_1\rightarrow0.999$) represents the fundamental material bound of bulk lithium niobate (LN), where only the intrinsic material attenuation ($\sim0.1$~dB/m) remains once extrinsic scattering losses are completely eliminated \cite{10.1063/5.0141436}.

Transitioning toward this idealized regime demands simultaneously overcoming the trade-offs between high parametric gain and waveguide insertion loss. Although cascading low-loss periodically poled lithium niobate (PPLN) waveguides provides a viable pathway for scaling, as discussed in our multi-stage architecture, achieving high gain within a single monolithic amplifier stage is preferable because it avoids the coupling losses and spatial alignment constraints associated with inter-stage connections.

Thin-film lithium niobate (TFLN) emerges as a highly promising hardware platform for realizing such high-gain, single-stage architectures. While our noise evaluation conservatively assumes a net PSA gain of $G_1\eta_1=30$~dB, the frontier of TFLN research has demonstrated remarkable milestones, albeit along largely separate research vectors. On one hand, high-gain parametric devices have achieved ultra-high gains exceeding 50~dB/cm \cite{Jankowski:22,Ledezma:22}. On the other hand, distinct optimization efforts focusing on structural quality have independently achieved ultra-low propagation losses of 2.7~dB/m \cite{Zhang:17}, closely approaching the theoretical fabrication limit of 0.2~dB/m \cite{10.1063/5.0095146}. Unifying these parallel advancements into a single, co-designed 1~cm device presents a crucial next engineering challenge; however, successfully combining both capabilities would yield a 50~dB gain with an internal propagation loss of merely 0.62\%, effectively reproducing the near-lossless conditions of our idealized regime. At the fundamental TFLN loss limit, this internal propagation loss would drop further to 0.046\%.

For the multi-stage architectures, the required secondary PIA gain of up to $G_2\approx17$~dB (see Table~\ref{tab:all_architecture_comparison_summary}) is already well within the mature performance envelope of existing semiconductor or fiber-based integrated linear optical amplifiers \cite{s23177326}, confirming the immediate technical viability of the hybrid setup in terms of the required gain. Note, however, that our analysis assumes a quantum-limited PIA (see Appendix~\ref{Evolution of Covariance Matrices under Quantum Operations}); since practical amplifiers exhibit noise figures above this limit, the noise performance of a near-term hybrid system would be correspondingly degraded compared with the values in Table~\ref{tab:all_architecture_comparison_summary}.

\subsection{Practical Feasibility of High-Speed Real-Time Operation}

The feasibility of high-speed real-time operation is evaluated across two distinct physical scales: the localized analog bandwidth of individual components and the macroscopic latency and throughput dynamics of the integrated network. For the latter, we distinguish the aggregate throughput of temporal modes from the clock rate of an individual computational line, i.e., the rate at which gate operations are sequentially applied to a single stream. Because the proposed gain-compensated architecture integrates PSAs, phase rotators, beam splitters, and secondary optical amplifiers (OAs), we first establish the localized bandwidth benchmarks for these constituent components.

Waveguide-based parametric amplifiers provide a localized processing bandwidth on the order of several THz \cite{Yamashima2025AllOpticalFeedforward, suzuki2026ultrafastallopticalquantumteleportation}, meaning that the operational speed is primarily bottlenecked by the phase switching rate of the rotators, which determines how fast the gate settings can be reprogrammed across successive time slots. Reprogramming the gate settings also requires switching the PSA pump phases ($-2\theta_k$) and the feedforward gain of Eq.~\eqref{required-gain}, which depends on $|\sin(\theta_2-\theta_1)|$. Both can be implemented with the same phase-control elements: the pump phase is switched directly by a phase rotator on the second-harmonic pump, and the gain is adjusted through the pump power by a Mach--Zehnder-type attenuator on the pump path, whose transmission is controlled by an internal phase rotator. Alternatively, the gain can be fixed at its maximum value and trimmed by a variable attenuator after the PSA, whose added noise is suppressed by the preceding high gain. The rate at which the gate settings are reprogrammed is therefore determined by the switching rate of phase rotators. To bypass these modulation bottlenecks and realize high-speed phase rotation, several broadband electro-optic and all-optical alternatives exist. TFLN-based EOMs offer direct electronic modulation bandwidths exceeding \SI{100}{GHz}~\cite{Wang20182, Yang:22, Valdez2022}. To push phase-control operations into the THz regime, all-optical switching based on the effectively instantaneous Kerr response---implemented through cross-phase modulation (XPM)---provides a viable pathway~\cite{Kupchak:19, sgjn-12vf, 974167}. Furthermore, nonlinear temporal focusing mirrors can compress chirped optical pulses into picosecond waveforms \cite{Berti2022}, enabling these phase-switching frameworks to sustain ultrafast operational cycles.

The remaining constituent components---specifically beam splitters and secondary OAs---similarly support comparable high-speed operation. Beam splitter operations are governed by instantaneous linear scattering, which is inherently broadband. Finally, secondary gain compensation can be implemented using THz-bandwidth PSAs or semiconductor optical amplifiers (SOAs). SOA gain can be optically switched through carrier-induced gain depletion and recovery, enabling ultrafast all-optical operation beyond the nominal carrier-recovery timescale \cite{Manning:97}; 100-Gb/s SOA-based all-optical logic has also been experimentally demonstrated \cite{Chen:16}.

While individual components can support localized analog bandwidths from hundreds of gigahertz to several terahertz and accommodate exceptionally high processing throughput, the integrated system as a whole remains bound by macroscopic temporal constraints. In the AOFF architecture, where optoelectronic conversions are completely eliminated, this system-level bottleneck is shifted from classical electronic processing delays to physical optical path lengths. The physical latency from measurement to displacement is strictly determined by the total transit time through the interconnected optical channels. Assuming a baseline configuration with a $45\text{-mm}$ PSA waveguide and a $100\text{-mm}$ highly nonlinear device for phase modulation \cite{Kupchak:19}, the total internal propagation latency is approximately $1\text{~ns}$.

When applying this feedforward architecture to TDM-MBQC, this $1\text{~ns}$ latency dictates the minimum feedforward loop time for an entangled pulse train, and thus bounds the clock rate of an individual computational line. Crucially, however, this propagation delay does not restrict the overall system throughput. While latency defines the transit time of an individual wavepacket, throughput depends on the capacity to continuously process ultrashort pulses at multi-GHz or THz repetition rates without mutual pulse interference. Consequently, any processing bottlenecks imposed by the $1\text{~ns}$ delay on a single TDM computational line can be mitigated by time-interleaving multiple independent quantum state streams within the latency window.

While this interleaving approach scales the aggregate throughput, the intrinsic clock rate of an individual computational line can be further accelerated by strategically executing the displacement operation downstream of the measurement apparatus. By applying the correction directly to the post-measurement state, the physical propagation delay incurred within the measurement stage is completely excluded from the real-time feedforward timing budget. This configuration handles the entire feedforward process all-optically, serving as an ultrafast analog to the paradigm of post-measurement information processing~\cite{yokoyama2026fullstackanalogopticalquantum}. Furthermore, because the primary PSA stage provides high parametric gain, the system's susceptibility to subsequent optical losses is significantly suppressed~\cite{PhysRevLett.119.223604, Inoue2023UltraFastOpticalQuantumProcessor, Kawasaki2024BroadbandNonGaussian, Kawasaki2025PicosecondEntanglement, Li:19, Shaked2018, PhysRevA.101.053801}. This attenuation tolerance enables the implementation of the remaining feedforward circuitry on integrated photonic platforms, which shortens the physical path lengths, minimizes propagation latency, and supports scalable, ultrafast real-time processing.

\medskip
In summary, the close alignment between the theoretical noise evaluations, the physical viability of the device parameters on emerging photonic platforms, and the high-speed capability of the all-optical feedforward loop validates the physical feasibility of the proposed configurations. Rather than relying on idealized assumptions, these configurations provide an insightful architectural framework that carefully balances rigorous quantum noise limits with empirical hardware capabilities. Ultimately, this work establishes a high-throughput, noise-resilient foundation for deploying scalable, ultra-high-speed building blocks in future quantum computing networks.

\section{Conclusion}\label{conclusion}

We have proposed an all-optical generalized quantum teleportation architecture for continuous-variable quantum information processing, in which a single stage executes programmable single-mode Gaussian operations and two successive stages realize arbitrary single-mode linear operations, and evaluated its comprehensive noise performance. The proposed framework establishes a robustness against practical propagation losses while enabling high-speed, all-optical feedforward processing that circumvents electronic latency and throughput bottlenecks; the analog bandwidth and the aggregate throughput are thereby set by the optical components rather than by electronics.

Under realistic hardware conditions—characterized by a net PSA gain of $G_1\eta_1=30$~dB with an internal efficiency of 93\%~\cite{suzuki2026ultrafastallopticalquantumteleportation,10.1063/5.0063118,10.1063/5.0144385}—the PSA-based (single-stage and dual-PSA) configurations maintain the maximum equivalent input noise below $-12$~dB for target squeezing operations up to 5~dB, which covers the Clifford operations required for GKP qubits (at most 4.18~dB), even in the presence of severe downstream propagation losses, while deeper target squeezing and the PSA-PIA configuration lead to higher noise levels. Furthermore, our analysis indicates that further minimizing the internal propagation loss of the PSA device can drastically suppress this noise floor toward near-zero levels. This resilience highlights the physical feasibility of loss-compensated all-optical feedforward operations, supporting its potential compatibility with the rigorous requirements of GKP-based FTQC. Additionally, for operational regimes demanding high target gains, the dual-PSA configuration successfully distributes the amplification burden without inducing a severe noise penalty.

Beyond single-mode operations, the proposed architecture inherently scales to larger multi-mode networks. Since quantum computation on TDM-MBQC platforms is fundamentally driven by gate sequences constructed solely from quantum teleportations and linear optical elements~\cite{PhysRevA.93.062326, yokoyama2026fullstackanalogopticalquantum}, this scheme implements the feedforward routines of these gate sequences entirely in the optical domain without optoelectronic conversion, demonstrating its high architectural flexibility.

In conclusion, by carefully balancing physical noise limitations with empirical hardware capabilities, this work delivers a high-throughput, noise-resilient structural foundation for next-generation ultrafast optical quantum processors. The synergy of loss robustness and all-optical feedforward speed establishes a practical framework for scaling continuous-variable quantum information processing.

\begin{acknowledgments}
This work was supported by JST [Moonshot R\&D][Grant No.~JPMJMS256I]. 
This paper is based on results obtained from a project, JPNP20017, subsidized by the New Energy and Industrial Technology Development Organization (NEDO).

The authors disclose financial competing interests: Takaya Hoshi, Akito Kawasaki, Xiruo Yan, Atsushi Sakaguchi, Takumi Suzuki, Tatsuki Sonoyama, Hironari Nagayoshi, and Kosuke Fukui are employees of OptQC Corp., and Kan Takase and Warit Asavanant are directors of, and hold equity in, OptQC Corp. In addition, OptQC Corp. has filed a patent application related to the work described in this manuscript.
\end{acknowledgments}

\appendix
\onecolumngrid

\section{Derivation of Generalized Quantum Teleportation}\label{Derivation of generalized quantum teleportation}

This appendix provides the derivation of Eq.~(\ref{output-GQT}) from the initial measurement relations in Eq.~(\ref{bell-measurement1}). We invoke the matrix inversion identity:
\begin{equation}
    \left[ \frac{1}{\sqrt{2}}\begin{pmatrix}
    -\sin\theta_1 & -\cos\theta_1 \\
    \sin\theta_2 & \cos\theta_2
    \end{pmatrix} \right]^{-1} = -\begin{pmatrix}
-1 & 0 \\
0 & 1
\end{pmatrix} \cdot \frac{\sqrt{2}}{\sin (\theta_2-\theta_1)} \begin{pmatrix}
\cos \theta_2 & \cos \theta_1 \\
\sin \theta_2 & \sin \theta_1
\end{pmatrix}.
\end{equation}
Isolating the input and ancillary quadrature terms in Eq.~(\ref{bell-measurement1}) yields:
\begin{equation}
    \begin{aligned}
        -\frac{\sqrt{2}}{\sin (\theta_2-\theta_1)} \begin{pmatrix}
\cos \theta_2 & \cos \theta_1 \\
\sin \theta_2 & \sin \theta_1
\end{pmatrix} \begin{pmatrix}
    m_1 \\
    m_2
    \end{pmatrix} &= \frac{-1}{\sin (\theta_2-\theta_1)} \begin{pmatrix}
\cos \theta_2 & \cos \theta_1 \\
\sin \theta_2 & \sin \theta_1
\end{pmatrix} \begin{pmatrix}
        \sin\theta_1 & \cos\theta_1 \\
        \sin\theta_2 & \cos\theta_2
    \end{pmatrix} \begin{pmatrix}
    \hat{x}_{\text{in}} \\
    \hat{p}_{\text{in}}
    \end{pmatrix} \\
    &+ \begin{pmatrix}
    -\hat{x}_{\text{A}} \\
    \hat{p}_{\text{A}}
    \end{pmatrix}.
    \end{aligned}
\end{equation}
The output quadratures before feedforward correction are then expressed as:
\begin{equation}
    \begin{aligned}
    \begin{pmatrix}
    \hat{x}_{\text{B}} \\
    \hat{p}_{\text{B}}
    \end{pmatrix} &= \begin{pmatrix}
    \hat{x}_{\text{B}} - \hat{x}_{\text{A}} \\
    \hat{p}_{\text{B}} + \hat{p}_{\text{A}}
    \end{pmatrix} + \mathbf{V}(\theta_1, \theta_2) \begin{pmatrix}
    \hat{x}_{\text{in}} \\
    \hat{p}_{\text{in}}
    \end{pmatrix} + \frac{\sqrt{2}}{\sin (\theta_2-\theta_1)} \begin{pmatrix}
\cos \theta_2 & \cos \theta_1 \\
\sin \theta_2 & \sin \theta_1
\end{pmatrix} \begin{pmatrix}
    m_1 \\
    m_2
    \end{pmatrix}.
    \end{aligned}
\end{equation}
Applying classical feedforward to cancel the measurement-dependent displacement term directly yields Eq.~(\ref{output-GQT}). 

Under an ideally squeezed Einstein-Podolsky-Rosen (EPR) resource state, the first term on the right-hand side satisfies the scaling relation:
\begin{equation}
    \begin{pmatrix}
    \hat{x}_{\text{B}} - \hat{x}_{\text{A}} \\
    \hat{p}_{\text{B}} + \hat{p}_{\text{A}}
    \end{pmatrix} = \sqrt{2}e^{-r} \begin{pmatrix}
    \hat{x}_{\mathrm{vac}} \\
    \hat{p}_{\mathrm{vac}}
    \end{pmatrix},
\end{equation}
which introduces an excess noise variance equivalent to twice that of the squeezed vacuum. The transfer matrix $\mathbf{V}(\theta_1, \theta_2)$ and its corresponding quantum operation $\hat{V}(\theta_1, \theta_2)$ are expressed as:
\begin{equation}\label{GQT-operation}
    \begin{aligned}
        \mathbf{V}(\theta_1, \theta_2) &= \frac{-1}{\sin (\theta_2-\theta_1)} \begin{pmatrix}
        \cos \theta_2 & \cos \theta_1 \\
        \sin \theta_2 & \sin \theta_1
        \end{pmatrix} \begin{pmatrix}
        \sin\theta_1 & \cos\theta_1 \\
        \sin\theta_2 & \cos\theta_2
        \end{pmatrix}, \\
        \hat{V}(\theta_1, \theta_2) &= \hat{R}\bigl(\pi\mathbf{H}(\theta_2-\theta_1)\bigr) \hat{R}\left(\frac{\theta_1+\theta_2}{2}-\frac{\pi}{2}\right) \hat{S}\left(-\log \left[\tan \left|\frac{\theta_1-\theta_2}{2}\right|\right]\right) \hat{R}\left(\frac{\theta_1+\theta_2}{2}\right),
    \end{aligned}
\end{equation}
where they satisfy the Heisenberg-picture transformation relation $\hat{V}^\dagger(\theta_1, \theta_2) \hat{\boldsymbol{r}} \hat{V}(\theta_1, \theta_2) = \mathbf{V}(\theta_1, \theta_2) \hat{\boldsymbol{r}}$ for the quadrature vector $\hat{\boldsymbol{r}} = (\hat{x}, \hat{p})^T$. Here, the first factor $\hat{R}(\pi\mathbf{H}(\theta_2-\theta_1))$ is the identity for $\theta_1>\theta_2$ and a $\pi$ rotation, whose action on $\hat{\boldsymbol{r}}$ is $\hat{\boldsymbol{r}}\to-\hat{\boldsymbol{r}}$, for $\theta_1<\theta_2$. It accounts for the overall sign of $\mathbf{V}(\theta_1,\theta_2)$ in the latter case, which cannot be represented by a scalar prefactor of the unitary operator because such a prefactor changes only the global phase. Here and throughout, the measurement angles are taken in the range $0<|\theta_1-\theta_2|<\pi$, and the numerical examples in this paper satisfy $\theta_1>\theta_2$, for which this factor is the identity.

This confirms that the operation induced by a single generalized quantum teleportation circuit consists of rotations and squeezing. For example, specific parameters yield:
\begin{equation}
\hat{V}(\theta_1=\arctan (e^{-r}), \theta_2=-\arctan (e^{-r})) = \hat{R}(-\pi/2) \hat{S}(r).
\end{equation}

Equation~\eqref{GQT-operation} also shows that a single GQT stage independently controls only one rotation angle, $(\theta_1+\theta_2)/2$, and the squeezing parameter, determined by $|\theta_1-\theta_2|$; the two rotations in Eq.~\eqref{GQT-operation} are not independent. An arbitrary single-mode linear transformation $\hat{R}(a)\hat{S}(s)\hat{R}(b)$ additionally requires one more programmable rotation, which is provided by adding another GQT stage configured as a pure rotation. For instance, choosing $\theta_1-\theta_2=\pi/2$ yields no squeezing, $\hat{V}(\beta+\pi/4,\beta-\pi/4)=\hat{R}(2\beta-\pi/2)$, with an arbitrary angle set by $\beta$. Therefore, two successive GQT stages, each implemented by the AOFF architecture proposed in this work, realize an arbitrary single-mode linear transformation.

\section{Decomposition and Phase Control of Phase-Sensitive Amplification}
\label{Decomposition and Phase Control of Phase-Sensitive Amplification}

This appendix provides the explicit derivation of the phase-controlled phase-sensitive amplifier (PSA) operator $\hat{S}_{\theta}(r)$, omitting the stage subscript $k$ for simplicity. 

\subsection{Physical Derivation from the Parametric Hamiltonian}
In a phase-matched single-mode optical parametric amplification process, the interaction Hamiltonian is given by:
\begin{equation}
    \hat{H} = \frac{g}{2} (\hat{a}_p^{\dagger} \hat{a}_s^2 + \hat{a}_p \hat{a}_s^{\dagger 2}),
\end{equation}
where $\hat{a}_p$ and $\hat{a}_s$ are the annihilation operators for the pump (second-harmonic) and signal (fundamental) modes, respectively, and $g$ is the nonlinear coupling constant. Under the parametric approximation where the pump is treated as an undepleted classical field ($\hat{a}_p = -i\beta = -i\gamma e^{i\phi_{\mathrm{pump}}}$), the time-evolution operator expands as:
\begin{align}
    \hat{U} = \exp( -i\hat{H}t ) 
    &= \exp \left\{ \frac{gt}{2}( \beta^*\hat{a}_s^2 - \beta\hat{a}_s^{\dagger2} ) \right\} \nonumber \\
    &= \hat{R}^\dagger\left( -\frac{\phi_{\mathrm{pump}}}{2} \right)\exp \left\{ \frac{gt}{2}( \gamma\hat{a}_s^2 - \gamma\hat{a}_s^{\dagger2} ) \right\}\hat{R}\left( -\frac{\phi_{\mathrm{pump}}}{2} \right) \nonumber \\
    &= \hat{R}^\dagger\left( -\frac{\phi_{\mathrm{pump}}}{2} \right)\hat{S}\left( gt\gamma \right)\hat{R}\left( -\frac{\phi_{\mathrm{pump}}}{2} \right).
\end{align}
Mapping the phase rotation angle to $\theta = -\phi_{\mathrm{pump}}/2$ establishes that a classical pump phase of $\phi_{\mathrm{pump}} = -2\theta$ yields the phase-controlled PSA operator:
\begin{equation}\label{app:psa_sandwich_def}
    \hat{S}_{\theta}(r) = \hat{R}^\dagger(\theta)\hat{S}(r)\hat{R}(\theta).
\end{equation}

\subsection{Algebraic Decomposition and Quadrature Transformation}
The adjoint transformation of the signal annihilation operator $\hat{a}$ under $\hat{S}_{\theta}(r)$ is evaluated as:
\begin{align}
    \hat{S}_{\theta}^\dagger\hat{a}\hat{S}_{\theta} 
    &= \hat{R}^\dagger(\theta)\left[ \hat{S}^\dagger(r)\left\{ \hat{R}(\theta)\hat{a}\hat{R}^\dagger(\theta) \right\}\hat{S}(r) \right]\hat{R}(\theta) \nonumber \\
    &= e^{-i\theta}\hat{R}^\dagger(\theta)\left\{\hat{S}^\dagger(r)\hat{a}\hat{S}(r) \right\}\hat{R}(\theta) \nonumber \\
    &= \hat{R}^\dagger(\theta)\left\{\frac{e^{-i\theta}}{\sqrt{2}} \left(\frac{1}{\sqrt{G}}\hat{x} + i\sqrt{G}\hat{p} \right)\right\}\hat{R}(\theta),
\end{align}
where $G = e^{2r}$ represents the operational power gain. Transforming the canonical quadratures into the rotated phase frame via $\hat{R}^\dagger(\theta)\hat{x}\hat{R}(\theta) = \hat{x}(\theta)$ and $\hat{R}^\dagger(\theta)\hat{p}\hat{R}(\theta) = \hat{p}(\theta)$ yields:
\begin{equation}\label{app:exact_psa_output}
    \hat{S}_{\theta}^\dagger\hat{a}\hat{S}_{\theta} = \frac{e^{-i\theta}}{\sqrt{2}} \left(\frac{1}{\sqrt{G}}\hat{x}(\theta) + i\sqrt{G}\hat{p}(\theta)\right).
\end{equation}

Equation~(\ref{app:exact_psa_output}) shows that the quadrature component $\hat{p}(\theta)$ is amplified by $\sqrt{G}$ while the orthogonal component $\hat{x}(\theta)$ is attenuated by $1/\sqrt{G}$.

In the high-gain regime ($G \gg 1$), the de-amplified term becomes negligible ($1/\sqrt{G} \to 0$; see Appendix~\ref{Residual Noise Contributions from Non-Amplified Quadratures} for the exact treatment without this approximation), allowing the asymptotic approximation:
\begin{align}
    \hat{S}_{\theta}^\dagger\hat{a}\hat{S}_{\theta} 
    &\approx \frac{e^{-i\theta}}{\sqrt{2}} \left( i\sqrt{G}\hat{p}(\theta) \right) \nonumber \\
    &= \frac{e^{i(\pi/2-\theta)}}{\sqrt{2}}\sqrt{G}\hat{m},
\end{align}
where $\hat{m} = \hat{p}(\theta)$ denotes the target quadrature component isolated by the PSA process.

\section{Exact Input--Output Relation beyond the High-Gain Approximation}\label{Residual Noise Contributions from Non-Amplified Quadratures}

In this appendix, we derive the complete Heisenberg input--output relation of the AOFF circuit of
Fig.~\ref{fig3} without the high-gain approximation of Eq.~\eqref{stepi-AOLFF}. To isolate this effect, all optical
components are assumed to be lossless ($\eta_1=1$, $L_1=0$); the PSA of Step~(i) is then described
exactly by Eq.~\eqref{eq:measop}, and all other elements are passive. We restrict the measurement
angles to $0<|\theta_1-\theta_2|<\pi$, so that the phase factor of Eq.~\eqref{ideal-feedfoward} can be
written as:
\begin{equation}
\sigma \equiv e^{i\pi\mathbf{H}(\theta_2-\theta_1)}.
\label{eq:C_sigma}
\end{equation}

\subsection{Feedforward field without the high-gain approximation}

Keeping the de-amplified term of Eq.~\eqref{eq:measop}, the output of the PSA in path $k$ reads exactly:
\begin{equation}
\hat{a}_{k,\text{out}}
=\frac{e^{i\left(\pi/2 - \theta_k\right)}}{\sqrt2}\left(\sqrt G\,\hat m_k - \frac{i}{\sqrt G}\,\hat x_k(\theta_k)\right).
\label{eq:C_akout}
\end{equation}
Applying the rotation of Step~(ii), including the $\pi$ phase shift associated with $\mathbf{H}(\theta_2-\theta_1)$,
Eq.~\eqref{stepii-AOLFF} is replaced by:
\begin{equation}
\hat a_1'=\sigma\frac{e^{i\theta_2}}{\sqrt2}\left(\sqrt G\,\hat m_1-\frac{i}{\sqrt G}\hat x_1(\theta_1)\right),\qquad
\hat a_2'=\sigma\frac{e^{i\theta_1}}{\sqrt2}\left(\sqrt G\,\hat m_2-\frac{i}{\sqrt G}\hat x_2(\theta_2)\right).
\label{eq:C_ak}
\end{equation}
The balanced beam splitter of Step~(iii) then yields, with $-i=e^{-i\pi/2}$:
\begin{equation}
\hat{P}_{\text{FF,AO,full}} = \frac{1}{\sqrt2}\left(\hat a_1'+\hat a_2'\right)
= \sigma\frac{\sqrt{G}}{2} \left( \hat{m}_1 e^{i\theta_2} + \hat{m}_2 e^{i\theta_1} \right)
+ \sigma\frac{1}{2\sqrt{G}}\left( \hat{x}_1(\theta_1)e^{i\left( \theta_2 - \pi/2 \right)} + \hat{x}_2(\theta_2)e^{i\left( \theta_1 - \pi/2 \right)} \right).
\label{eq:C_PFF}
\end{equation}
The first term reproduces Eq.~\eqref{measinfo-AO}, including the $\pi$ phase shift depending on the measurement bases. The second term is the
contribution neglected in Eq.~\eqref{stepi-AOLFF}; its direction in phase space is rotated by $-\pi/2$
relative to the amplified signal. Defining the quadrature operators of the feedforward field as
$\hat{\mathcal{X}}_{\text{FF}}\equiv(\hat{P}_{\text{FF,AO,full}}+\hat{P}_{\text{FF,AO,full}}^\dagger)/\sqrt2$ and
$\hat{\mathcal{P}}_{\text{FF}}\equiv(\hat{P}_{\text{FF,AO,full}}-\hat{P}_{\text{FF,AO,full}}^\dagger)/(i\sqrt2)$,
and taking $\sqrt2$ times the real and imaginary parts of the complex coefficients in Eq.~\eqref{eq:C_PFF}, we obtain:
\begin{equation}
\begin{pmatrix}\hat{\mathcal{X}}_{\text{FF}}\\ \hat{\mathcal{P}}_{\text{FF}}\end{pmatrix}
=\sigma\sqrt{\frac{G}{2}}\begin{pmatrix}\cos\theta_2&\cos\theta_1\\ \sin\theta_2&\sin\theta_1\end{pmatrix}
\begin{pmatrix}\hat m_1\\ \hat m_2\end{pmatrix}
+\frac{\sigma}{\sqrt{2G}}\begin{pmatrix}\sin\theta_2&\sin\theta_1\\ -\cos\theta_2&-\cos\theta_1\end{pmatrix}
\begin{pmatrix}\hat x_1(\theta_1)\\ \hat x_2(\theta_2)\end{pmatrix}.
\label{eq:C_FF}
\end{equation}
The first term has the same form as the measurement-dependent term of Eq.~\eqref{output-GQT}, whereas the
second term originates from the de-amplified quadratures.

\subsection{Output mode after the displacement}

The feedforward field of Eq.~\eqref{eq:C_FF} is coupled to mode $\hat a_B$ at the displacement beam splitter
as in Eq.~\eqref{output-GQT}, where $\sqrt{1-T}(\hat{\mathcal{X}}_{\text{FF}},\hat{\mathcal{P}}_{\text{FF}})^{\mathrm T}$
now replaces both the classical displacement and the vacuum contribution. As in Sec.~\ref{Operational Principle and Derivation}, the first term of Eq.~\eqref{eq:C_FF} cancels the measurement-dependent term of mode $B$ exactly under the gain condition of Eq.~\eqref{required-gain}. The output mode therefore consists of the first two terms in the brackets of Eq.~\eqref{output-GQT}, multiplied by $\sqrt T$, and the contribution of the second term of Eq.~\eqref{eq:C_FF}, which reads, using Eq.~\eqref{required-gain}:
\begin{equation}
-\frac{\sqrt{2T}}{G\sin(\theta_2-\theta_1)}
\begin{pmatrix}\sin\theta_2&\sin\theta_1\\ -\cos\theta_2&-\cos\theta_1\end{pmatrix}
\begin{pmatrix}\hat x_1(\theta_1)\\ \hat x_2(\theta_2)\end{pmatrix}.
\label{eq:C_pre}
\end{equation}
The de-amplified quadratures are expressed in terms of the input and resource modes in the same manner as
Eq.~\eqref{bell-measurement1}, using $\hat x_k(\theta_k)=\hat x_k\cos\theta_k-\hat p_k\sin\theta_k$:
\begin{equation}
\begin{pmatrix}\hat x_1(\theta_1)\\ \hat x_2(\theta_2)\end{pmatrix}
=\frac{1}{\sqrt2}\begin{pmatrix}\cos\theta_1&-\sin\theta_1\\ \cos\theta_2&-\sin\theta_2\end{pmatrix}
\begin{pmatrix}\hat x_{\text{in}}\\ \hat p_{\text{in}}\end{pmatrix}
+\frac{1}{\sqrt2}\begin{pmatrix}-\cos\theta_1&\sin\theta_1\\ \cos\theta_2&-\sin\theta_2\end{pmatrix}
\begin{pmatrix}\hat x_A\\ \hat p_A\end{pmatrix}.
\label{eq:C_xi}
\end{equation}
They thus contain both the input and one half of the EPR resource, and cannot be regarded as independent
vacuum noise. Substituting Eq.~\eqref{eq:C_xi} into Eq.~\eqref{eq:C_pre}, the input part becomes:
\begin{equation}
\frac{\sqrt T}{G}\cdot\frac{-1}{\sin(\theta_2-\theta_1)}
\begin{pmatrix}\sin(\theta_1+\theta_2)&-2\sin\theta_1\sin\theta_2\\ -2\cos\theta_1\cos\theta_2&\sin(\theta_1+\theta_2)\end{pmatrix}
\begin{pmatrix}\hat x_{\text{in}}\\ \hat p_{\text{in}}\end{pmatrix}
=\frac{\sqrt T}{G}\,\mathbf V^{-{\rm T}}(\theta_1,\theta_2)\begin{pmatrix}\hat x_{\text{in}}\\ \hat p_{\text{in}}\end{pmatrix},
\end{equation}
where the inverse transpose of $\mathbf V(\theta_1,\theta_2)$ in Eq.~\eqref{GQT-operation} has been used. Similarly, the resource part becomes:
\begin{equation}
\frac{\sqrt T}{G}\cdot\frac{-1}{\sin(\theta_2-\theta_1)}
\begin{pmatrix}-\sin(\theta_2-\theta_1)&0\\ 0&\sin(\theta_2-\theta_1)\end{pmatrix}
\begin{pmatrix}\hat x_A\\ \hat p_A\end{pmatrix}
=\frac{\sqrt T}{G}\begin{pmatrix}\hat x_A\\ -\hat p_A\end{pmatrix}.
\end{equation}
We thus obtain:
\begin{equation}
\boxed{
\begin{pmatrix}\hat x_{\text{out}}\\ \hat p_{\text{out}}\end{pmatrix}
=\sqrt T\Bigl(\mathbf V+\frac{1}{G}\mathbf V^{-{\rm T}}\Bigr)\begin{pmatrix}\hat x_{\text{in}}\\ \hat p_{\text{in}}\end{pmatrix}
+\sqrt T\begin{pmatrix}\hat x_B-\hat x_A\\ \hat p_B+\hat p_A\end{pmatrix}
+\frac{\sqrt T}{G}\begin{pmatrix}\hat x_A\\ -\hat p_A\end{pmatrix}}
\label{eq:C_result}
\end{equation}
with $G$ given by Eq.~\eqref{required-gain}. Equation~\eqref{eq:C_result} is exact.
Eq.~\eqref{output-GQT} with the feedforward of Eq.~\eqref{ideal-feedfoward} is recovered in the limit $T\to1$,
where $G\to\infty$ by Eq.~\eqref{required-gain}. The finite-gain and finite-$T$ effects thus reduce to two contributions:
a deterministic modification of the input transformation (first term), and terms
originating from the EPR resource (second and third terms).
Note that no additional vacuum mode enters Eq.~\eqref{eq:C_result}. The feedforward field consists of the amplified quadratures $\hat m_k$ and the de-amplified quadratures $\hat x_k(\theta_k)$, which are canonically conjugate to each other and thus ensure the commutation relation of the feedforward mode. In the electronic scheme of Eq.~\eqref{output-GQT}, this role is played by the vacuum fluctuations accompanying the feedforward field, so that the factor $\sqrt T$ is equivalent to an optical loss of $1-T$; in the AOFF scheme, the factor $\sqrt T$ is instead compensated by the de-amplified quadratures and does not act as an optical loss.
Although lossless components have been assumed here, the same derivation applies to the lossy PSA and the propagation loss considered in Sec.~\ref{sec:Performance_Analysis}: with $G$ in Eq.~\eqref{required-gain} understood as the effective gain $G_1\eta_1(1-L_1)$, the deterministic parts of Eq.~\eqref{eq:C_result} retain the same form with $1/G$ replaced by the ratio of the de-amplified to the amplified amplitude transmission, namely $1/G_1$ for the single-stage and PSA-PIA configurations and $1/(G_1G_2)$ for the dual-PSA configuration, both of which do not exceed $1/G$, whereas the vacuum fluctuations injected by the losses give the additive noise covariance evaluated in Sec.~\ref{sec:Performance_Analysis}, so that the two contributions can be treated independently.

\subsection{Deterministic gate transformation}

We now show that the first term of Eq.~\eqref{eq:C_result} belongs to the same family of operations as
Eq.~\eqref{GQT-operation}. Following the relation between $\hat V(\theta_1,\theta_2)$ and
$\mathbf V(\theta_1,\theta_2)$ in Eq.~\eqref{GQT-operation}, we denote by $\mathbf R(\phi)$ and $\mathbf S(r)$
the matrices describing the Heisenberg-picture actions of $\hat R(\phi)$ and $\hat S(r)$ on
$\hat{\boldsymbol r}=(\hat x,\hat p)^{\rm T}$. From Eqs.~\eqref{eq:rotation-op} and \eqref{eq:squeezing-op}, they are:
\begin{equation}
\mathbf R(\phi)=\begin{pmatrix}\cos\phi&-\sin\phi\\ \sin\phi&\cos\phi\end{pmatrix},\qquad
\mathbf S(r)=\begin{pmatrix}e^{-r}&0\\0&e^{r}\end{pmatrix}.
\end{equation}
Since $\mathbf R(\pi\mathbf H(\theta_2-\theta_1))=\sigma\,\mathbf I$ with $\sigma$ of Eq.~\eqref{eq:C_sigma},
the decomposition in Eq.~\eqref{GQT-operation} reads:
\begin{equation}
\mathbf V(\theta_1,\theta_2)=\sigma\,\mathbf R(\alpha)\,\mathbf S(r)\,\mathbf R(\beta),\qquad
\alpha=\frac{\theta_1+\theta_2}{2}-\frac{\pi}{2},\quad \beta=\frac{\theta_1+\theta_2}{2},
\end{equation}
where the gate squeezing $r=-\ln|\tan[(\theta_1-\theta_2)/2]|$ is related to the measurement bases by $|\sin(\theta_1-\theta_2)|=1/\cosh r$.
Because $\mathbf R(\phi)$ is orthogonal and $\mathbf S(r)^{-{\rm T}}=\mathbf S(-r)$, we have
$\mathbf V^{-{\rm T}}(\theta_1,\theta_2)=\sigma\,\mathbf R(\alpha)\,\mathbf S(-r)\,\mathbf R(\beta)$; namely,
$\mathbf V^{-{\rm T}}$ corresponds to the operation of Eq.~\eqref{GQT-operation} with the squeezing reversed. Therefore:
\begin{equation}
\sqrt T\Bigl(\mathbf V+\frac{1}{G}\mathbf V^{-{\rm T}}\Bigr)
=\sqrt T\,\sigma\,\mathbf R(\alpha)\Bigl(\mathbf S(r)+\frac{1}{G}\mathbf S(-r)\Bigr)\mathbf R(\beta)
=\sqrt T\lambda\;\sigma\,\mathbf R(\alpha)\,\mathbf S(r')\,\mathbf R(\beta)
=\sqrt T\lambda\;\mathbf V(\theta_1',\theta_2'),
\label{eq:C_gate}
\end{equation}
where $\lambda$ and $r'$ are determined by $e^{-r}+e^{r}/G=\lambda e^{-r'}$ and $e^{r}+e^{-r}/G=\lambda e^{r'}$:
\begin{align}
\lambda&=\sqrt{\Bigl(1+\frac{e^{2r}}{G}\Bigr)\Bigl(1+\frac{e^{-2r}}{G}\Bigr)},
&\sqrt T\lambda&\simeq1-\frac{1-T}{4\cosh^2 r},\\
r'&=r-\frac12\ln\frac{1+e^{2r}/G}{1+e^{-2r}/G},
&r'-r&\simeq-\frac{1-T}{2T}\tanh r,
\label{eq:C_rprime}
\end{align}
where the approximate forms are first order in $1-T$, obtained from Eq.~\eqref{required-gain} with $|\sin(\theta_1-\theta_2)|=1/\cosh r$.
Here $\theta_1'+\theta_2'=\theta_1+\theta_2$, $|\tan[(\theta_1'-\theta_2')/2]|=e^{-r'}$, and
$\theta_1'-\theta_2'$ has the same sign as $\theta_1-\theta_2$, so that $\alpha$, $\beta$, and $\sigma$ are unchanged.
The input thus undergoes the same family of Gaussian operations $\mathbf V(\theta_1',\theta_2')$:
the rotation angles are unchanged, the squeezing is slightly reduced ($r\to r'$), and the
output is rescaled by the isotropic factor $\sqrt T\lambda$. No input-dependent noise
arises. Because the squeezing modification $r\to r'$ is a known function of $(\theta_1,\theta_2,T)$, it can
be precompensated by retuning $\theta_1-\theta_2$. In contrast, the isotropic factor $\sqrt T\lambda$ cannot be
removed by recalibrating the feedforward gain, since the gain is fixed by the cancellation condition of
Eq.~\eqref{required-gain}; its effect is estimated in Sec.~\ref{sec:Discussion}.

\subsection{Resource-induced terms}

The second term of Eq.~\eqref{eq:C_result} is the usual EPR noise of Eq.~\eqref{output-GQT}, rescaled by
$\sqrt T$; it vanishes in the ideal-resource limit. The third term couples to an individual
EPR quadrature and does not vanish in that limit. It is independent of the input and
amounts to an imbalance of the EPR cancellation. For a two-mode squeezed resource with
squeezing parameter $r_E$, $\langle\hat x_A^2\rangle=\langle\hat p_A^2\rangle=\cosh(2r_E)/2$,
so this term contributes the variance:
\begin{equation}
\frac{T\cosh(2r_E)}{2G^2}=\frac{(1-T)^2\cosh(2r_E)}{32\,T\cosh^4 r}
\end{equation}
per quadrature, where $r$ is the gate squeezing introduced above. In the following, we characterize the EPR resource by its squeezing level $10\log_{10}e^{2r_E}$, i.e., the squeezing level of the two single-mode squeezed states from which it is generated. Since this contribution grows as $\cosh(2r_E)$, we evaluate it at high resource squeezing levels of 25 and 30~dB to obtain a pessimistic estimate. Relative to the vacuum
level, it is at most $-30.0$ and $-25.0$~dB, respectively, for $T=0.99$, and at most $-50.0$ and
$-45.0$~dB for $T=0.999$ (Table~\ref{tab:C}). Since this term is correlated with the second term through
$\hat x_A$ and $\hat p_A$, its net effect on the output variance is
$T[\cosh(2r_E)/(2G^2)-e^{-2r_E}/G]$ per quadrature, which is bounded above by the variance of the term
itself; the correlation reduces the output variance only for moderate resource squeezing, whereas for
resource squeezing levels of 25 and 30~dB the net effect is positive and close to the variance of the term. Because the contribution is
isotropic, it adds to the EIN as $e^{2r}$ and $e^{-2r}$ times its variance along the squeezed and
anti-squeezed axes of the gate, respectively. For $T=0.99$, this increases the maximum EIN of the configurations in
Table~\ref{tab:all_architecture_comparison_summary} by at most $0.16$~dB for a resource squeezing level of 25~dB and
$0.5$~dB for 30~dB, whereas for $T=0.999$ the increase is below $0.2$~dB even for 30~dB. Since the contribution scales as $(1-T)^2\cosh(2r_E)$, a transmittance $T$ closer to unity
becomes important when highly squeezed resources are employed. The impact on the GKP-based discussion is
evaluated in Sec.~\ref{sec:Discussion}.

\begin{table}[h]
\caption{Exact finite-gain modification, Eqs.~\eqref{eq:C_gate}--\eqref{eq:C_rprime}, for the squeezing
gates of Sec.~\ref{sec:Performance_Analysis}. $G$ follows from Eq.~\eqref{required-gain}. $\Delta S=20\log_{10}e\,(r'-r)$ is the change in
gate squeezing. The last two columns are the variance of the resource term of Eq.~\eqref{eq:C_result}
relative to the vacuum level, for resource squeezing levels of 25 and 30~dB.}
\label{tab:C}
\begin{ruledtabular}
\begin{tabular}{cccccccc}
 & & & & & & \multicolumn{2}{c}{resource term (dB)}\\
$T$ & $S$ (dB) & $G$ (dB) & $\Delta S$ (dB) & $\sqrt T\lambda$ & $\sqrt T$ & 25-dB resource & 30-dB resource\\
\hline
0.99 & 0 & 26.0 & $0$ & 0.99750 & 0.99499 & $-30.0$ & $-25.0$\\
0.99 & 5 & 27.3 & $-0.023$ & 0.99817 & 0.99499 & $-32.7$ & $-27.7$\\
0.99 & 10 & 30.8 & $-0.036$ & 0.99917 & 0.99499 & $-39.6$ & $-34.6$\\
0.999 & 0 & 36.0 & $0$ & 0.99975 & 0.99950 & $-50.0$ & $-45.0$\\
0.999 & 5 & 37.4 & $-0.002$ & 0.99982 & 0.99950 & $-52.8$ & $-47.8$\\
0.999 & 10 & 40.8 & $-0.004$ & 0.99992 & 0.99950 & $-59.7$ & $-54.7$\\
\end{tabular}
\end{ruledtabular}
\end{table}

\subsection{Sensitivity to calibration errors}

The derivation above, as well as the covariance analysis in Sec.~\ref{sec:Performance_Analysis}, assumes that the PSA pump phases, the common rotation $\varphi$, the gain calibration of Eq.~\eqref{required-gain}, and the beam-splitter ratios are set exactly. If any of these parameters deviates from its nominal value, the feedforward field no longer cancels the Bell-dependent term of mode $B$ exactly, and a residual term linear in the quadratures of the Bell modes remains. Since these quadratures contain only the input and mode $\hat a_A$ of the EPR resource [Eqs.~\eqref{bell-measurement1} and \eqref{eq:C_xi}], the residual term has two effects: (i) a coherent error of the input transfer matrix, such as a rescaling and quadrature cross-talk, which is not captured by the additive EIN, and (ii) a leakage of the quadratures of mode $\hat a_A$, whose variance $\cosh(2r_E)/2$ is large for a highly squeezed resource. A gain error or a beam-splitter imbalance changes the amplitude of the feedforward field, and an error of the common rotation $\varphi$ rotates the feedforward field in phase space. A misalignment of the PSA axis has two effects: the PSA extracts a slightly different quadrature, so that part of the orthogonal quadrature that should be discarded leaks into the feedforward signal, and the amplified field is emitted in a slightly rotated direction in phase space. The same structure appears in the conventional electronic feedforward, whereas the AOFF architecture additionally requires the PSA pump phases to be locked.

To illustrate these effects quantitatively, we consider a gain error $\epsilon$, by which the amplitude of the whole feedforward field deviates by a factor of $1+\epsilon$ from the value calibrated by Eq.~\eqref{required-gain}, e.g., owing to a deviation of the parametric gain, $G\to(1+\epsilon)^2G$. We denote the Bell-dependent term of Eq.~\eqref{output-GQT}, i.e., the third term in the brackets, by $\hat{\boldsymbol b}$. Since the decomposition of mode $B$ in Appendix~\ref{Derivation of generalized quantum teleportation} is an identity for $(\hat x_B,\hat p_B)^{\rm T}$, this term is directly expressed by the input and resource modes as:
\begin{equation}
\hat{\boldsymbol b}=\begin{pmatrix}\hat x_A\\ -\hat p_A\end{pmatrix}-\mathbf V(\theta_1,\theta_2)\begin{pmatrix}\hat x_{\text{in}}\\ \hat p_{\text{in}}\end{pmatrix}.
\label{eq:C_bell_identity}
\end{equation}
With the calibrated gain, the feedforward field contributes $-\sqrt T\,\hat{\boldsymbol b}$ to the output, which cancels the term $\sqrt T\,\hat{\boldsymbol b}$ of mode $B$. The gain error changes this contribution to $-(1+\epsilon)\sqrt T\,\hat{\boldsymbol b}$, so that the Bell-dependent term is over-cancelled by:
\begin{equation}
\delta\begin{pmatrix}\hat x_{\text{out}}\\ \hat p_{\text{out}}\end{pmatrix}
\simeq-\epsilon\sqrt T\,\hat{\boldsymbol b}
=\epsilon\sqrt T\left[\mathbf V\begin{pmatrix}\hat x_{\text{in}}\\ \hat p_{\text{in}}\end{pmatrix}-\begin{pmatrix}\hat x_A\\ -\hat p_A\end{pmatrix}\right],
\label{eq:C_calib}
\end{equation}
where corrections of order $\epsilon/G$ arising from the de-amplified part of the feedforward field are neglected. The first term rescales the target transformation by $1+\epsilon$, and the second term is the leakage of mode $\hat a_A$. Since the leaked quadratures are correlated with the resource-dependent terms of Eq.~\eqref{eq:C_result}, $\sqrt T(\hat x_B-\hat x_A,\hat p_B+\hat p_A)^{\rm T}$ and $(\sqrt T/G)(\hat x_A,-\hat p_A)^{\rm T}$, the change of the output variance per quadrature relative to the vacuum level is:
\begin{equation}
\Delta V\simeq T\left[\epsilon^2\cosh(2r_E)-\frac{2\epsilon\cosh(2r_E)}{G}+2\epsilon e^{-2r_E}\right].
\label{eq:C_calib_var}
\end{equation}
Here, the first term is the variance of the leaked quadratures themselves, and the second and third terms are their cross terms with $(\sqrt T/G)(\hat x_A,-\hat p_A)^{\rm T}$ and $\sqrt T(\hat x_B-\hat x_A,\hat p_B+\hat p_A)^{\rm T}$, respectively. The cross terms, linear in $\epsilon$, make the added noise depend on the sign of the error: a slight over-amplification ($\epsilon>0$) partially compensates the under-cancellation of mode $\hat a_A$ by the fraction $1/G$ in Eq.~\eqref{eq:C_result}, whereas an under-amplification ($\epsilon<0$) increases the leakage.

For $T=0.99$ and target squeezing of 0 and 5~dB, a gain error of $\pm1\%$ yields a relative error of the input transfer matrix of $1\%$, and an added noise of up to $-16.3$~dB for a resource squeezing level of 25~dB and $-11.3$~dB for 30~dB, the latter being comparable to the hardware-induced maximum EIN of Table~\ref{tab:all_architecture_comparison_summary}. Keeping this added noise below $-20$~dB requires the gain to be calibrated within approximately $0.6\%$ and $0.3\%$, respectively. The other errors, such as phase errors in the PSA axis or the common rotation and an imbalance of the balanced beam splitter, produce residual terms of the same structure and are therefore expected to lead to added noise of a similar order for errors of comparable size. Highly squeezed resources therefore require correspondingly accurate calibration of the feedforward, as is also the case for the conventional electronic feedforward.

\section{Evolution of Covariance Matrices under Quantum Operations}\label{Evolution of Covariance Matrices under Quantum Operations}

This appendix summarizes the transformation rules for covariance matrices under each fundamental operational block.

\subsection{Optical Loss and Pure Rotation}
An optical loss $L$ is modeled as a beam splitter coupling the signal mode to a vacuum environment, yielding the transformation matrix $W_{\text{loss}}(L)$ and the additive noise covariance matrix $V_{\xi, \text{loss}}(L)$:
\begin{equation}
    W_{\text{loss}}(L) = \sqrt{1-L} \cdot \mathbf{I}, \quad V_{\xi, \text{loss}}(L) = \frac{1}{2} L \cdot \mathbf{I}.
\end{equation}
A phase shift $\phi$ is represented as a noiseless unitary rotation:
\begin{equation}
    W_{\text{Rot}}(\phi) = \begin{pmatrix} \cos\phi & -\sin\phi \\ \sin\phi & \cos\phi \end{pmatrix}, \quad V_{\xi, \text{Rot}} = \mathbf{0}.
\end{equation}

\subsection{PSA with Insertion Loss}
A phase-controlled phase-sensitive amplifier (PSA) with operational gain $G$ and internal efficiency $\eta$ amplifies along the quadrature $\hat{p}(\theta)$. The transformation is decomposed into a core squeezing operator $W_{\text{sq}}$ sandwiched by rotation matrices $W_{\text{Rot}}$:
\begin{equation}
    W_{\text{PSA}}(\theta, G, \eta) = W_{\text{Rot}}(-\theta) W_{\text{sq}}(G, \eta) W_{\text{Rot}}(\theta), \quad V_{\xi,\text{PSA}}(\theta, G, \eta) = W_{\text{Rot}}(-\theta)V_{\xi,\text{sq}}(G, \eta)W^T_{\text{Rot}}(-\theta),
\end{equation}
where the canonical squeezing transformation and its associated noise covariance are:
\begin{equation}
    W_{\text{sq}}(G, \eta) = \begin{pmatrix} \sqrt{\frac{\eta}{G}} & 0 \\ 0 & \sqrt{G\eta} \end{pmatrix}, \quad V_{\xi, \text{sq}}(G, \eta) = \frac{1}{2} \begin{pmatrix} N'_{\text{PSA}} & 0 \\ 0 & N_{\text{PSA}} \end{pmatrix}.
\end{equation}
The internal noise parameters are defined as $N_{\text{PSA}} = \frac{\ln \eta}{\ln \eta + \ln G}(1-G\eta)$ and $N'_{\text{PSA}} = \frac{\ln \eta}{\ln \eta - \ln G}(1-\frac{\eta}{G})$. In the high-gain limit ($G \gg 1$), the noise is dominated by the amplified quadrature, rendering $N'_{\text{PSA}}$ negligible (the full physical derivation is provided in Appendix~\ref{Physical Modeling of PSA with Internal Propagation Loss}).

\subsection{Phase-Insensitive Amplifier (PIA)}
As an alternative to the dual-PSA setup for multi-stage configurations, a phase-insensitive amplifier (PIA) amplifies both canonical quadratures isotropically. To preserve the quantum commutation relations, a PIA inevitably introduces excess vacuum noise. For an ideal, quantum-limited PIA with power gain $G$, the transformation matrix $W_{\text{PIA}}$ and additive noise covariance $V_{\xi, \text{PIA}}$ are expressed as:
\begin{equation}
    W_{\text{PIA}}(G) = \sqrt{G} \cdot \mathbf{I}, \quad V_{\xi, \text{PIA}}(G) = \frac{G-1}{2} \cdot \mathbf{I}.
\end{equation}
The noise term $(G-1)/2$ represents the minimum quantum noise required to maintain the uncertainty principle bound under isotropic amplification, corresponding to the operator-level relation:
\begin{equation}
    \hat{a}_{\text{out}} = \sqrt{G} \hat{a}_{\text{in}} + \sqrt{G-1} \hat{a}_{\text{vac}}^{\dagger},
\end{equation}
where $\hat{a}_{\text{vac}}^{\dagger}$ is the creation operator of the vacuum field coupled through the unused port. Any excess technical noise is neglected to evaluate the fundamental performance limit of the architecture.

\subsection{Beam Splitter Operation}
The interference of two spatial modes at a beam splitter with power transmittance $T$ transforms the joint quadrature vector $\hat{\boldsymbol{r}}_{\text{in}} = (\hat{x}_1, \hat{p}_1, \hat{x}_2, \hat{p}_2)^T$ via the $4 \times 4$ matrix:
\begin{equation}\label{BS-covariance-transformation}
    W_{\text{BS}}(T) = \begin{pmatrix} 
        \sqrt{1-T} \cdot \mathbf{I} & -\sqrt{T} \cdot \mathbf{I} \\ 
        \sqrt{T} \cdot \mathbf{I} & \sqrt{1-T} \cdot \mathbf{I} 
    \end{pmatrix}, \quad V_{\xi, \text{BS}} = \mathbf{0},
\end{equation}
where $\mathbf{I}$ denotes the $2 \times 2$ identity matrix. This framework models both the balanced beam splitter ($T=1/2$) and the asymmetric displacement port ($T \to 1$). As a passive, lossless component, the beam splitter introduces no additional noise ($V_{\xi, \text{BS}} = \mathbf{0}$).

\section{Physical Modeling of PSA with Internal Propagation Loss}\label{Physical Modeling of PSA with Internal Propagation Loss}

Following the continuous-loss model established in previous work~\cite{Yamashima2025AllOpticalFeedforward}, we consider a lossy PSA characterized by a parametric gain $G = e^{2r} \neq 1$ and an internal transmission efficiency $\eta \leq 1$. To model the simultaneous coexistence of amplification (or de-amplification) and distributed internal loss, the process is discretized into $n$ infinitesimal stages. For each iterative step, the quadrature operators evolve as:
\begin{equation}
\begin{aligned}
    \hat{x}_{k+1} &= \left( \eta^{\frac{1}{2n}} \hat{x}_k + \sqrt{1-\eta^{\frac{1}{n}}} \hat{x}_{\text{vac}} \right) e^{-r/n}, \\
    \hat{p}_{k+1} &= \left( \eta^{\frac{1}{2n}} \hat{p}_k + \sqrt{1-\eta^{\frac{1}{n}}} \hat{p}_{\text{vac}} \right) e^{r/n}.
\end{aligned}
\end{equation}

By integrating these simultaneous dynamics and taking the continuum limit $n \rightarrow \infty$ in the iterative process, the final output covariance matrix $V_{\text{out}}$ is analytically derived as:
\begin{equation}\label{cov-lossyPSA}
    V_{\text{out}} = W_{\text{PSA}} V_{\text{in}} W_{\text{PSA}}^T + V_{\xi, \text{PSA}},
\end{equation}
where the deterministic transformation matrix is $W_{\text{PSA}} = \sqrt{\eta} \,\text{diag}(1/\sqrt{G}, \sqrt{G})$ and the injected excess noise matrix is $V_{\xi, \text{PSA}} = \text{diag}(V_{\xi, x}, V_{\xi, p})$. The diagonal components of this injected noise are given by:
\begin{equation} \label{PSA-noise-term}
\begin{aligned}
    V_{\xi, x} &= \frac{1}{2} \frac{\ln \eta}{\ln \eta - \ln G} \left( 1 - \frac{\eta}{G} \right), \\
    V_{\xi, p} &= \frac{1}{2} \frac{\ln \eta}{\ln \eta + \ln G} (1 - G\eta).
\end{aligned}
\end{equation}

Note that while these expressions formally encounter $0/0$ indeterminate forms at two specific parameter critical points, they evaluate to well-defined, continuous physical values via L'Hôpital's rule. Specifically, at the amplification critical point where $G\eta = 1$ ($G > 1$), the noise component converges to $\lim_{G\eta \to 1} V_{\xi, p} = \frac{1}{2}\ln G$. Symmetrically, at the de-amplification critical point where $\eta/G = 1$ ($G < 1$), the other component scales smoothly to $\lim_{\eta \to G} V_{\xi, x} = -\frac{1}{2}\ln G$. For the ideal lossless limit ($\eta = 1$), both excess noise terms cleanly vanish ($V_{\xi, x} = V_{\xi, p} = 0$). This ensures that the formulated quantum noise remains strictly bounded, non-singular, and physically continuous across the entire operational domain of $G \neq 1$ and $\eta \leq 1$.

\section{Derivation of the Overall Noise Covariance Matrices}\label{app:noise-derivation}

This appendix details the algebraic derivation of the accumulated noise covariance matrices for the AOFF configurations analyzed in Section~\ref{sec:Performance_Analysis}.

\subsection{General Noise Synthesis and Feedforward Output}
In both the single-stage and multi-stage configurations, the extracted measurement signals from each path are combined at a balanced beam splitter (Step~iii) before driving the displacement on mode~$\hat{a}_B$. According to Eq.~\eqref{general-noise-sum}, the synthesized noise covariance matrix $V_{\text{FF}}$ at the feedforward output port (denoted by the subscript index $22$) is related to the individual path noise matrices $V_1$ and $V_2$ by:
\begin{equation}
    V_{\text{FF}} = \left[ W_{\text{BS}}\left( \frac{1}{2} \right) \begin{pmatrix} V_1 & \mathbf{0} \\ \mathbf{0} & V_2 \end{pmatrix} W_{\text{BS}}^T\left( \frac{1}{2} \right) \right]_{22} = \frac{1}{2}(V_1 + V_2).
\end{equation}
To isolate the noise contribution of the AOFF circuit, mode~$\hat{a}_B$ is assumed to be initially in a noise-free state. The total output noise covariance matrix $V_{\text{out}}$ after coupling through the final beam splitter with power transmittance $T$ is expressed as:
\begin{equation}
    V_{\text{out}} = \left[ W_{\text{BS}}(T) \begin{pmatrix} V_{\text{FF}} & \mathbf{0} \\ \mathbf{0} & \mathbf{0} \end{pmatrix} W_{\text{BS}}^T(T) \right]_{11} = (1-T) V_{\text{FF}} = \frac{1-T}{2}(V_1 + V_2).
\end{equation}
Substituting the architecture-specific formulations of $V_k$ ($k=1,2$) into this general relation yields the final output noise profiles.

\subsection{Derivation of Path Noise for the Single-Stage Architecture}
For the single-stage building block, the signal in each path $k$ sequentially undergoes phase-sensitive amplification, a propagation loss $L_1$, and a phase rotation $\varphi$. To maintain notational brevity, explicit functional arguments of the transformation matrices and noise vectors are suppressed here. Applying the elementary transformation rules from Appendix~\ref{Evolution of Covariance Matrices under Quantum Operations}, the intermediate covariance matrix immediately following the propagation loss $L_1$ is given by:
\begin{equation}
    V_{k,\text{loss}} = (1-L_1) V_{\xi, \text{PSA}, k} + V_{\xi, \text{loss1}}.
\end{equation}
Applying the final phase rotation $W_{\text{Rot}}$ yields the complete path noise covariance matrix $V_k$:
\begin{equation}
    V_k = W_{\text{Rot}} \left[ (1-L_1) V_{\xi, \text{PSA}, k} + V_{\xi, \text{loss1}} \right] W_{\text{Rot}}^T.
\end{equation}
Combining $V_k$ with the general synthesis relation defines the total output noise covariance matrix for the single-stage architecture:
\begin{equation}
    V_{\text{out}} = \frac{1-T}{2} \sum_{k=1}^{2} W_{\text{Rot}} \left[ (1-L_1) V_{\xi, \text{PSA}, k} + V_{\xi, \text{loss1}} \right] W_{\text{Rot}}^T,
\end{equation}
where $V_{\xi, \text{PSA}, k}$ represents the intrinsic excess noise of the PSA in path $k$ (parameterized by the phase axis $\theta_k$, gain $G_1$, and internal efficiency $\eta_1$), $V_{\xi, \text{loss1}}$ denotes the vacuum fluctuations injected via the consolidated external loss $L_1$, and $W_{\text{Rot}}$ is the phase rotation matrix $W_{\text{Rot}}(\varphi)$.

\subsection{Derivation of Path Noise for the Multi-Stage Architecture}
For the multi-stage configurations, each path incorporates a secondary optical amplifier (OA) and a subsequent propagation loss $L_2$ prior to the phase rotation. To maintain consistency, explicit functional arguments are again suppressed throughout this derivation. Using the intermediate noise covariance matrix $V_{\text{out}, k}$ accumulated after the first-stage loss $L_1$ defined in Eq.~\eqref{inter-stage-noise}, the covariance matrix immediately following the general OA stage transforms as:
\begin{equation}
    V_{k,\text{OA}} = W_{\text{OA}} V_{\text{out}, k} W_{\text{OA}}^T + V_{\xi, \text{OA}}.
\end{equation}
The subsequent propagation loss $L_2$ scales this accumulated noise and injects vacuum fluctuations $V_{\xi, \text{loss2}}$, yielding:
\begin{equation}
    V_{k,\text{loss2}} = (1-L_2) V_{k,\text{OA}} + V_{\xi, \text{loss2}}.
\end{equation}
Applying the final phase rotation $W_{\text{Rot}}$ gives the complete multi-stage path noise matrix $V_k$:
\begin{equation}
    V_k = W_{\text{Rot}} \Bigl[ (1-L_2) \Bigl\{ W_{\text{OA}} V_{\text{out}, k} W_{\text{OA}}^T + V_{\xi, \text{OA}} \Bigr\} + V_{\xi, \text{loss2}} \Bigr] W_{\text{Rot}}^T.
\end{equation}
Substituting this path noise into the general synthesis relation yields the total output noise covariance matrix for the Dual-PSA configuration:
\begin{equation}
    V_{\text{out, multi}}^{\text{PSA}} = \frac{1-T}{2} \sum_{k=1}^{2} W_{\text{Rot}} \Bigl[ (1-L_2) \Bigl\{ W_{\text{PSA2}, k} V_{\text{out}, k} W_{\text{PSA2}, k}^T + V_{\xi, \text{PSA2}} \Bigr\} + V_{\xi, \text{loss2}} \Bigr] W_{\text{Rot}}^T.
\end{equation}
Similarly, replacing the second-stage PSA with a phase-insensitive amplifier (PIA) yields the output noise covariance matrix for the hybrid PSA-PIA architecture:
\begin{equation}
    V_{\text{out, multi}}^{\text{PIA}} = \frac{1-T}{2} \sum_{k=1}^{2} W_{\text{Rot}} \Bigl[ (1-L_2) \Bigl\{ W_{\text{PIA}} V_{\text{out}, k} W_{\text{PIA}}^T + V_{\xi, \text{PIA}} \Bigr\} + V_{\xi, \text{loss2}} \Bigr] W_{\text{Rot}}^T.
\end{equation}
Here, the general operators $W_{\text{OA}}$ and $V_{\xi, \text{OA}}$ are explicitly mapped to either the phase-sensitive parameters ($W_{\text{PSA2}}(\theta_k, G_2, \eta_2)$, $V_{\xi, \text{PSA2}}(\theta_k, G_2, \eta_2)$) or the phase-insensitive parameters ($W_{\text{PIA}}(G_2)$, $V_{\xi, \text{PIA}}(G_2)$) depending on the choice of the secondary architecture, while $V_{\xi, \text{loss2}}$ represents the vacuum fluctuations injected via the propagation loss $L_2$.

\subsection{Geometric Orientation of Squeezed Thermal Noise}

To clarify the directional behavior of the single-stage total noise, we analyze the geometric orientation of its noise ellipse. The covariance matrix of a non-ideal PSA can be structurally decomposed into a canonical, unrotated squeezed thermal noise matrix $V_{\xi, \text{sq}}$ sandwiched by phase rotation matrices:
\begin{equation}\label{eq:psa_rotation_decomposition}
    V_{\xi, \text{PSA}}(\theta_k, G_1, \eta_1) = W_{\text{Rot}}(-\theta_k) V_{\xi, \text{sq}}(G_1, \eta_1) W_{\text{Rot}}^T(-\theta_k).
\end{equation}
Substituting Eq.~\eqref{eq:psa_rotation_decomposition} into the total output noise formula and exploiting the linearity of matrix transformations, the isotropic loss covariance $V_{\xi, \text{loss}}(L_1) = \frac{L_1}{2}\mathbf{I}$ remains invariant under orthogonal rotations. The total noise covariance matrix $V_{\text{out}}$ can therefore be expressed by aggregating the directional actions of the phase rotations:
\begin{equation}
    V_{\text{out}} = \frac{1-T}{2} (1-L_1) \sum_{k=1}^{2} W_{\text{Rot}}(\varphi - \theta_k) V_{\xi, \text{sq}} W_{\text{Rot}}^T(\varphi - \theta_k) + \mathcal{V}_{\text{iso}},
\end{equation}
where $\mathcal{V}_{\text{iso}} = \frac{1-T}{2} L_1 \mathbf{I}$ represents the isotropic background noise component that does not contribute to directional skewness.

To track the principal orientation of the anisotropic noise component, we apply the operational phase-lock condition $\varphi = \theta_1 + \theta_2 - \pi/2$. Under this constraint, the net rotation angles simplify symmetrically to:
\begin{align}
    \varphi - \theta_1 &= \theta_2 - \frac{\pi}{2}, \\
    \varphi - \theta_2 &= \theta_1 - \frac{\pi}{2}.
\end{align}
Given the diagonal form of the intrinsic noise covariance matrix $V_{\xi, \text{sq}} = \operatorname{diag}(v_x, v_y)$, the summation of the two rotated noise terms can be simplified by eliminating the cross-terms, yielding a single rotation structure centered around a modified effective diagonal matrix $V_{\text{eff}}$:
\begin{equation}\label{eq:sum_rotated_noise_expanded}
    \sum_{k=1}^{2} W_{\text{Rot}}(\varphi - \theta_k) V_{\xi, \text{sq}} W_{\text{Rot}}^T(\varphi - \theta_k) = W_{\text{Rot}}\left(\frac{\theta_1 + \theta_2}{2}- \frac{\pi}{2}\right) V_{\text{eff}} W_{\text{Rot}}^T\left(\frac{\theta_1 + \theta_2}{2}- \frac{\pi}{2}\right),
\end{equation}
where the synthesized core noise matrix $V_{\text{eff}}$ preserves its strict diagonal form in the designated rotation basis and is analytically defined as:
\begin{equation}\label{eq:V_eff_definition}
    V_{\text{eff}} = \begin{pmatrix} (v_x+v_y) + (v_x-v_y)\cos(\theta_1-\theta_2) & 0 \\ 0 & (v_x+v_y) - (v_x-v_y)\cos(\theta_1-\theta_2) \end{pmatrix}.
\end{equation}
Substituting this representation back into the total output noise formula gives:
\begin{equation}
    V_{\text{out}} = \frac{1-T}{2}(1-L_1) W_{\text{Rot}}\left(\frac{\theta_1 + \theta_2}{2}- \frac{\pi}{2}\right) V_{\text{eff}} W_{\text{Rot}}^T\left(\frac{\theta_1 + \theta_2}{2}- \frac{\pi}{2}\right) + \mathcal{V}_{\text{iso}}.
\end{equation}

This analytical form determines the geometric orientation of the noise distribution relative to the induced quantum operation. Under parametric amplification, the anti-squeezing noise power dominates the squeezing noise power ($v_y > v_x$). Notably, the sign of $\cos(\theta_1-\theta_2)$ simultaneously dictates both the anti-squeezing axis of the target gate $\hat{V}(\theta_1, \theta_2)$ defined in Eq.~\eqref{GQT-operation} and the dominant noise quadrature in $V_{\text{eff}}$. When $\cos(\theta_1-\theta_2) > 0$, the gate operation anti-squeezes the local $p$-axis, which precisely coincides with the maximum noise variance of the $(2,2)$-element in $V_{\text{eff}}$. Conversely, when $\cos(\theta_1-\theta_2) < 0$, both the gate's anti-squeezing axis and the dominant noise variance simultaneously flip to the local $x$-axis (the $(1,1)$-element). Consequently, after mapping back to the laboratory frame via the outer rotation $W_{\text{Rot}}\left(\frac{\theta_1 + \theta_2}{2} - \frac{\pi}{2}\right)$, the major axis of the total synthesized noise ellipse remains deterministically and perfectly aligned with the exact anti-squeezing direction of the executed gate. This robust geometric synchronization remains invariant even in multi-stage configurations; the second-stage PSA induces additional scaling along the same phase axis, while the excess noise from the PIA and associated propagation losses are isotropic ($\mathcal{V}_{\text{iso}} \propto \mathbf{I}$), naturally preserving the primary alignment between the total noise ellipse and the quantum gate operation.

\twocolumngrid

\bibliography{main}

\end{document}